\documentclass[useAMS,usenatbib]{mn2e}
\usepackage{graphicx}
\title[Electron impact excitation of C III]{Energy levels, radiative rates and electron impact excitation rates for transitions in C III\thanks{Tables 2, 7 and 8 are available only in the electronic version.}}
\author[K. M. Aggarwal and F. P. Keenan]{Kanti  M.  ~Aggarwal$^{1}$\thanks{E-mail:
 K.Aggarwal@qub.ac.uk(KMA); F.Keenan@qub.ac.uk (FPK)} and Francis  P.   ~Keenan$^{1}$ \\
$^{1}$Astrophysics Research Centre, School of Mathematics and Physics, Queen's University Belfast, Belfast BT7 1NN, Northern Ireland, UK} 
\begin{document}

\date{Accepted 2015 March 25. Received 2015 March 12; in original form 2015 February 6}

\pagerange{\pageref{firstpage}--\pageref{lastpage}} \pubyear{2015}

\maketitle

\label{firstpage}

\begin{abstract}
We report energy levels,  radiative rates (A-values) and lifetimes for the astrophysically-important Be-like ion C III. For the calculations, 166 levels belonging to  the $n \le$ 5 configurations are considered and the {\sc grasp} (General-purpose Relativistic Atomic Structure Package) is adopted. Einstein A-coefficients  are provided  for all E1, E2, M1 and M2 transitions, while lifetimes are compared  with  available measurements as well as theoretical results,  and no large discrepancies  noted.  Our  energy levels are  assessed to be accurate to better than 1\% for a majority of levels, and A-values to better than 20\% for most transitions.   Collision strengths are also calculated, for which the Dirac Atomic R-matrix Code ({\sc darc}) is used.  A wide energy range, up to 21 Ryd, is considered  and resonances  resolved in a fine energy mesh in the thresholds region. The  collision strengths are subsequently averaged over a Maxwellian velocity distribution to determine effective collision strengths up to a  temperature of 8.0$\times$10$^5$ K, sufficient for most astrophysical applications. Our data are compared with the recent $R$-matrix calculations of   \cite*{icft}, and significant differences (up to over an order of magnitude) are noted for several transitions over the complete temperature range of the results.

\end{abstract}


\section{Introduction}

Observations of emission lines from Be-like ions, such as C~III, O~V, Ca~XVII and Fe~XXIII, provide useful  diagnostics for  density and temperature  of astrophysical plasmas  -- see for example, \cite{el1}.  Considering C~III, early observations of  strong emission ($\lambda$: 5696 $\rm \AA$) and  absorption  ($\lambda$: 4647 $\rm \AA$) features in HD~192639 were reported by \cite{ss40}. \cite{kw} deduced electron densities using  several line pairs arising from the  $n$ = 2 transitions of C~III in  solar spectra obtained by the Naval Research Laboratory's S082B instrument on board {\em Skylab}. More recently, strong  emission lines of C~III ($\lambda\lambda$ 4647-4650-4652 $\rm\AA$) have been detected in spectroscopic survey of Galactic O-type stars by \cite{nolan}. Similarly, among ultraviolet emission lines in young low-mass galaxies at redshift z $\sim$ 2, \cite{stark} noted that the strongest  (other than Ly$\alpha$) is always the blended C~III $\lambda$1908 doublet. Many  observed lines of C~III  below 4700 $\rm \AA$ are listed in the CHIANTI database at {\tt http://www.chiantidatabase.org/}. Similarly, numerous transitions  in the 270--2070 $\rm \AA$ wavelength range are included in  the {\em Atomic Line List} (v2.04) of Peter van Hoof at ${\tt {\verb+http://www.pa.uky.edu/~peter/atomic/+}}$. 

For the reliable interpretation of observations, accurate atomic data are required for several parameters, such as energy levels, Einstein coefficients or radiative rates (A-values) and   excitation rates -- see section 6. Energy levels for C~III have been calculated by several workers --  see for example, \cite{gu} and references therein. Measurements of energy levels have been compiled and critically evaluated by the  NIST (National Institute of Standards and Technology) team \citep{nist} and are available at their  website {\tt http://www.nist.gov/pml/data/asd.cfm}. Similarly, A-values have been reported by numerous workers -- see for example, \cite{uiss, uis}.

Since  C~III is an important ion in astrophysical (and fusion) plasmas, a few workers have reported data  for collision strengths $\Omega$, or more importantly {\em effective} collision strengths  $\Upsilon$, which are directly related to the excitation rates (section 6). Early $R$-matrix calculations were mostly performed by Berrington and his colleagues - see in particular  \cite{kab2} and references therein. These  included 12 states among $n \le$ 3 configurations and were in $LS$  coupling (Russell-Saunders or spin-orbit coupling). Subsequently, \cite{mit} extended the calculations to 24 states (including additional states of the $n$ = 4  configurations), and made  other improvements, particularly in the generation of wavefunctions.  However, their results for $\Upsilon$ have been shown to be overestimated (by up to 50\%, particularly at lower temperatures) for several important transitions, such as 2s$^2$ $^1$S$_0$ -- 2s2p $^3$P$^o_{1,2}$ and 2s$^2$ $^1$S$_0$ -- 2p$^2$ $^1$D$_2$ -- see fig. 3 of \cite*{icft}. The calculations of \cite{icft} are not only the most recent but also the most extensive, because they have reported data for transitions among 238 {\em fine-structure} levels, belonging to the $n \le$ 7 configurations of C~III. 

To determine energy levels and A-values,  \cite{icft}  adopted the {\em AutoStructure} (AS) code of \cite{as}. The  wave functions generated  were subsequently employed in  the $R$-matrix code of \cite*{rm2} for  the  calculations of $\Omega$, primarily obtained in  $LS$ coupling.  Corresponding results for desired  fine-structure transitions were  determined  through their intermediate coupling frame transformation (ICFT) method. Furthermore, they resolved resonances in the thresholds region to finally determine values of $\Upsilon$. Since C~III is a comparatively light ion, higher-order relativistic effects (neglected in the AS and $R$-matrix codes) are not very important, and hence their approach should give comparable results with a fully relativistic version, such as the Dirac atomic  $R$-matrix code (DARC). Unfortunately, this is not the case as recently demonstrated by us \citep{belike} for five other Be-like ions, namely Al~X, Cl~XIV, K~XVI, Ti~XIX  and Ge~XXIX. For about 50\% of the transitions (of all ions) the $\Upsilon$ of \cite{icft} are overestimated, up to more than an order of magnitude, and over a wide range of temperatures. Therefore, considering the importance of C~III, we decided  to perform yet another calculation so that atomic data can be confidently applied to the modelling of plasmas.

As in all our earlier work on Be-like ions, we have  employed the fully relativistic {\sc grasp} (General-purpose Relativistic Atomic Structure  Package) code,  originally  developed by  \cite{grasp0}, but  significantly revised by one of its authors (Dr. P. H. Norrington),   and  available at the website {\tt http://web.am.qub.ac.uk/DARC/}, to determine the atomic structure, i.e.  to calculate energy levels and A-values. Similarly for the scattering calculations, we have adopted the unpublished DARC code of P. H. Norrington and I. P. Grant,  freely available at {\tt http://web.am.qub.ac.uk/DARC/}). This is a relativistic version of the standard $R$-matrix code and is based on the $jj$ coupling scheme. 

\begin{table*}
 \centering
 \caption{Energy levels (in Ryd) of C III and their lifetimes (s).  ($a{\pm}b \equiv a{\times}$10$^{{\pm}b}$).}
\begin{tabular}{rllrrrrrl} \hline
Index  & Configuration       & Level          &  NIST      &   GRASP   &  AS     & $\tau$ (s) \\
\hline
    1  &    2s$^2$  &  $^1$S$  _0$   &  0.00000  &  0.00000 &  0.00000    &  .....      \\  
    2  &    2s2p    &  $^3$P$^o_0$   &  0.47720  &  0.48390 &  0.48949    &  .....      \\  
    3  &    2s2p    &  $^3$P$^o_1$   &  0.47742  &  0.48409 &  0.48981    &  1.578$-$02 \\
    4  &    2s2p    &  $^3$P$^o_2$   &  0.47793  &  0.48457 &  0.49045    &  1.854$+$02 \\
    5  &    2s2p    &  $^1$P$^o_1$   &  0.93270  &  1.00439 &  1.00281    &  4.648$-$10 \\
    6  &    2p$^2$  &  $^3$P$  _0$   &  1.25232  &  1.27817 &  1.29046    &  6.993$-$10 \\
    7  &    2p$^2$  &  $^3$P$  _1$   &  1.25258  &  1.27842 &  1.29077    &  6.990$-$10 \\
    8  &    2p$^2$  &  $^3$P$  _2$   &  1.25301  &  1.27880 &  1.29140    &  6.985$-$10 \\
    9  &    2p$^2$  &  $^1$D$  _2$   &  1.32932  &  1.39544 &  1.40723    &  7.464$-$09 \\
   10  &    2p$^2$  &  $^1$S$  _0$   &  1.66324  &  1.76957 &  1.77435    &  3.718$-$10 \\
   11  &    2s3s    &  $^3$S$  _1$   &  2.17076  &  2.15962 &  2.14181    &  2.635$-$10 \\
   12  &    2s3s    &  $^1$S$  _0$   &  2.25238  &  2.24672 &  2.23168    &  1.099$-$09 \\
   13  &    2s3p    &  $^3$P$^o_0$   &  2.36661  &  2.35629 &  2.34023    &  1.312$-$08 \\
   14  &    2s3p    &  $^3$P$^o_1$   &  2.36666  &  2.35632 &  2.34030    &  4.252$-$09 \\
   15  &    2s3p    &  $^3$P$^o_2$   &  2.36678  &  2.35646 &  2.34040    &  1.309$-$08 \\
   16  &    2s3p    &  $^1$P$^o_1$   &  2.35956  &  2.35674 &  2.33989    &  2.657$-$10 \\
   17  &    2s3d    &  $^3$D$  _1$   &  2.46052  &  2.45212 &  2.43547    &  9.623$-$11 \\
   18  &    2s3d    &  $^3$D$  _2$   &  2.46053  &  2.45213 &  2.43549    &  9.625$-$11 \\
   19  &    2s3d    &  $^3$D$  _3$   &  2.46056  &  2.45215 &  2.43553    &  9.628$-$11 \\
   20  &    2s3d    &  $^1$D$  _2$   &  2.51950  &  2.52902 &  2.50907    &  1.551$-$10 \\
   21  &    2p3s    &  $^3$P$^o_0$   &  2.80868  &  2.80571 &  2.79183    &  3.217$-$10 \\
   22  &    2p3s    &  $^3$P$^o_1$   &  2.80897  &  2.80600 &  2.79215    &  3.215$-$10 \\
   23  &    2p3s    &  $^3$P$^o_2$   &  2.80959  &  2.80660 &  2.79281    &  3.211$-$10 \\
   24  &    2s4s    &  $^3$S$  _1$   &  2.81998  &  2.80756 &  2.79138    &  3.790$-$10 \\
   25  &    2p3s    &  $^1$P$^o_1$   &  2.82499  &  2.82522 &  2.81029    &  2.406$-$10 \\
   26  &    2s4s    &  $^1$S$  _0$   &  2.84062  &  2.82683 &  2.81094    &  8.800$-$10 \\
   27  &    2s4p    &  $^3$P$^o_0$   &  2.89595  &  2.88219 &  2.86598    &  3.374$-$09 \\
   28  &    2s4p    &  $^3$P$^o_1$   &  2.89597  &  2.88220 &  2.86600    &  3.374$-$09 \\
   29  &    2s4p    &  $^3$P$^o_2$   &  2.89602  &  2.88225 &  2.86604    &  3.372$-$09 \\
   30  &    2p3p    &  $^1$P$  _1$   &  2.91351  &  2.91034 &  2.89504    &  2.971$-$10 \\
   31  &    2s4d    &  $^3$D$  _1$   &  2.92892  &  2.91830 &  2.90239    &  1.838$-$10 \\
   32  &    2s4d    &  $^3$D$  _2$   &  2.92906  &  2.91836 &  2.90245    &  1.840$-$10 \\
   33  &    2s4d    &  $^3$D$  _3$   &  2.92927  &  2.91846 &  2.90254    &  1.843$-$10 \\
   34  &    2s4f    &  $^3$F$^o_2$   &  2.93431  &  2.92159 &  2.90549    &  1.219$-$09 \\
   35  &    2s4f    &  $^3$F$^o_3$   &  2.93437  &  2.92163 &  2.90554    &  1.218$-$09 \\
   36  &    2s4f    &  $^3$F$^o_4$   &  2.93444  &  2.92169 &  2.90560    &  1.216$-$09 \\
   37  &    2s4f    &  $^1$F$^o_3$   &  2.94068  &  2.92696 &  2.91036    &  5.571$-$10 \\
   38  &    2p3p    &  $^3$D$  _1$   &  2.94409  &  2.94399 &  2.92673    &  1.039$-$08 \\
   39  &    2p3p    &  $^3$D$  _2$   &  2.94432  &  2.94428 &  2.92704    &  9.865$-$09 \\
   40  &    2p3p    &  $^3$D$  _3$   &  2.94467  &  2.94475 &  2.92753    &  9.157$-$09 \\
   41  &    2s4d    &  $^1$D$  _2$   &  2.95444  &  2.95066 &  2.93185    &  2.387$-$10 \\
   42  &    2s4p    &  $^1$P$^o_1$   &  2.93796  &  2.95239 &  2.93307    &  6.210$-$10 \\
   43  &    2p3p    &  $^3$S$  _1$   &  2.98238  &  2.98492 &  2.96582    &  4.977$-$10 \\
   44  &    2p3p    &  $^3$P$  _0$   &  3.00431  &  3.00370 &  2.99238    &  5.816$-$10 \\
   45  &    2p3p    &  $^3$P$  _1$   &  3.00451  &  3.00388 &  2.99259    &  5.815$-$10 \\
   46  &    2p3p    &  $^3$P$  _2$   &  3.00484  &  3.00419 &  2.99298    &  5.815$-$10 \\
   47  &    2p3d    &  $^1$D$^o_2$   &  3.03171  &  3.02604 &  3.01196    &  1.891$-$10 \\
   48  &    2p3d    &  $^3$F$^o_2$   &  3.03805  &  3.04299 &  3.02730    &  1.913$-$09 \\
   49  &    2p3d    &  $^3$F$^o_3$   &  3.03827  &  3.04322 &  3.02757    &  1.921$-$09 \\
   50  &    2p3d    &  $^3$F$^o_4$   &  3.03859  &  3.04355 &  3.02795    &  1.931$-$09 \\
\hline
\end{tabular} 
\end{table*}

\setcounter{table}{0}
\begin{table*}
\caption{... continued}
\begin{tabular}{rllrrrrrl} \hline
Index  & Configuration       & Level          &  NIST      &   GRASP   &  AS     & $\tau$ (s) \\
\hline           
   51  &    2p3p    &  $^1$D$  _2$   &  3.03560 &  3.05501 &  3.04207    &  4.404$-$10 \\
   52  &    2p3d    &  $^3$D$^o_1$   &  3.07695 &  3.07184 &  3.05766    &  8.004$-$11 \\
   53  &    2p3d    &  $^3$D$^o_2$   &  3.07707 &  3.07195 &  3.05779    &  8.005$-$11 \\
   54  &    2p3d    &  $^3$D$^o_3$   &  3.07724 &  3.07212 &  3.05798    &  8.005$-$11 \\
   55  &    2s5s    &  $^1$S$  _0$   &  3.08477 &  3.07632 &  3.06083    &  1.564$-$09 \\
   56  &    2s5s    &  $^3$S$  _1$   &  3.09771 &  3.08391 &  3.06670    &  1.150$-$09 \\
   57  &    2p3d    &  $^3$P$^o_2$   &  3.09924 &  3.09682 &  3.08003    &  2.223$-$10 \\
   58  &    2p3d    &  $^3$P$^o_1$   &  3.09947 &  3.09699 &  3.08021    &  2.238$-$10 \\
   59  &    2p3d    &  $^3$P$^o_0$   &  3.09960 &  3.09708 &  3.08030    &  2.246$-$10 \\
   60  &    2s5f    &  $^1$F$^o_3$   &  3.11080 &  3.11518 &  3.09884    &  2.444$-$10 \\
   61  &    2s5p    &  $^1$P$^o_1$   &  3.12800 &  3.12205 &  3.10437    &  3.236$-$10 \\
   62  &    2s5p    &  $^3$P$^o_2$   &  3.13688 &  3.12668 &  3.10904    &  3.918$-$10 \\
   63  &    2s5p    &  $^3$P$^o_1$   &  3.13691 &  3.12677 &  3.10912    &  3.863$-$10 \\
   64  &    2s5p    &  $^3$P$^o_0$   &  3.13693 &  3.12683 &  3.10916    &  3.834$-$10 \\
   65  &    2s5d    &  $^3$D$  _1$   &  3.14840 &  3.13429 &  3.11775    &  3.908$-$10 \\
   66  &    2s5d    &  $^3$D$  _2$   &  3.14840 &  3.13430 &  3.11776    &  3.910$-$10 \\
   67  &    2s5d    &  $^3$D$  _3$   &  3.14840 &  3.13430 &  3.11776    &  3.912$-$10 \\
   68  &    2s5g    &  $^3$G$  _3$   &  3.15826 &  3.14245 &  3.12578    &  2.899$-$09 \\
   69  &    2s5g    &  $^3$G$  _4$   &  3.15826 &  3.14245 &  3.12578    &  2.899$-$09 \\
   70  &    2s5g    &  $^3$G$  _5$   &  3.15826 &  3.14245 &  3.12578    &  2.899$-$09 \\
   71  &    2s5g    &  $^1$G$  _4$   &  3.15826 &  3.14246 &  3.12579    &  2.899$-$09 \\
   72  &    2s5f    &  $^3$F$^o_2$   &  3.16348 &  3.15099 &  3.13397    &  1.078$-$09 \\
   73  &    2s5f    &  $^3$F$^o_3$   &  3.16349 &  3.15101 &  3.13400    &  1.077$-$09 \\
   74  &    2s5f    &  $^3$F$^o_4$   &  3.16351 &  3.15105 &  3.13403    &  1.076$-$09 \\
   75  &    2s5d    &  $^1$D$  _2$   &  3.15898 &  3.15210 &  3.13247    &  3.905$-$10 \\
   76  &    2p3d    &  $^1$P$^o_1$   &  3.15948 &  3.17556 &  3.15654    &  1.547$-$10 \\
   77  &    2p3p    &  $^1$S$  _0$   &  3.14474 &  3.18667 &  3.15528    &  8.363$-$10 \\
   78  &    2p3d    &  $^1$F$^o_3$   &  3.17905 &  3.18957 &  3.17320    &  1.309$-$10 \\
   79  &    2p4s    &  $^3$P$^o_0$   &  3.42890 &  3.42158 &  3.40783    &  5.587$-$10 \\
   80  &    2p4s    &  $^3$P$^o_1$   &  3.42909 &  3.42188 &  3.40815    &  5.580$-$10 \\
   81  &    2p4s    &  $^3$P$^o_2$   &  3.43004 &  3.42250 &  3.40881    &  5.565$-$10 \\
   82  &    2p4s    &  $^1$P$^o_1$   &  3.47289 &  3.45238 &  3.44104    &  4.086$-$10 \\
   83  &    2p4p    &  $^1$P$  _1$   &          &  3.46581 &  3.45272    &  3.562$-$10 \\
   84  &    2p4p    &  $^3$D$  _1$   &  3.48058 &  3.47424 &  3.46093    &  8.093$-$10 \\
   85  &    2p4p    &  $^3$D$  _2$   &  3.48078 &  3.47457 &  3.46127    &  8.105$-$10 \\
   86  &    2p4p    &  $^3$D$  _3$   &  3.48113 &  3.47508 &  3.46180    &  8.103$-$10 \\
   87  &    2p4p    &  $^3$S$  _1$   &          &  3.49002 &  3.47643    &  4.567$-$10 \\
   88  &    2p4p    &  $^3$P$  _0$   &  3.50241 &  3.49717 &  3.48323    &  8.461$-$10 \\
   89  &    2p4p    &  $^3$P$  _1$   &  3.50259 &  3.49736 &  3.48343    &  8.444$-$10 \\
   90  &    2p4p    &  $^3$P$  _2$   &  3.50296 &  3.49766 &  3.48377    &  8.458$-$10 \\
   91  &    2p4d    &  $^1$D$^o_2$   &  3.51583 &  3.50794 &  3.49477    &  3.984$-$10 \\
   92  &    2p4d    &  $^3$F$^o_2$   &  3.51591 &  3.50991 &  3.49692    &  1.214$-$09 \\
   93  &    2p4d    &  $^3$F$^o_3$   &  3.51591 &  3.51017 &  3.49721    &  1.280$-$09 \\
   94  &    2p4d    &  $^3$F$^o_4$   &  3.51591 &  3.51060 &  3.49767    &  1.283$-$09 \\
   95  &    2p4p    &  $^1$D$  _2$   &  3.51419 &  3.51943 &  3.50408    &  5.997$-$10 \\
   96  &    2p4d    &  $^3$D$^o_1$   &  3.53296 &  3.52476 &  3.51043    &  1.661$-$10 \\
   97  &    2p4d    &  $^3$D$^o_2$   &  3.53296 &  3.52487 &  3.51055    &  1.662$-$10 \\
   98  &    2p4d    &  $^3$D$^o_3$   &  3.53296 &  3.52504 &  3.51074    &  1.661$-$10 \\
   99  &    2p4f    &  $^1$F$  _3$   &  3.54276 &  3.52790 &  3.51381    &  6.067$-$10 \\
  100  &    2p4f    &  $^3$F$  _2$   &  3.53728 &  3.52842 &  3.51444    &  6.357$-$10 \\
\hline
\end{tabular} 
\end{table*}

\setcounter{table}{0}
\begin{table*}
\caption{... continued}
\begin{tabular}{rllrrrrrl} \hline
Index  & Configuration       & Level          &  NIST      &   GRASP   &  AS     & $\tau$ (s) \\
\hline 
  101  &    2p4f    &  $^3$F$  _3$   &  3.53739 &  3.52851 &  3.51453    &  6.351$-$10 \\
  102  &    2p4f    &  $^3$F$  _4$   &  3.53767 &  3.52860 &  3.51464    &  6.357$-$10 \\
  103  &    2p4d    &  $^3$P$^o_2$   &  3.54021 &  3.53408 &  3.51984    &  2.603$-$10 \\
  104  &    2p4d    &  $^3$P$^o_1$   &  3.54021 &  3.53437 &  3.52013    &  2.600$-$10 \\
  105  &    2p4d    &  $^3$P$^o_0$   &  3.54021 &  3.53452 &  3.52028    &  2.599$-$10 \\
  106  &    2p4f    &  $^3$G$  _3$   &  3.53686 &  3.53684 &  3.52268    &  7.018$-$10 \\
  107  &    2p4f    &  $^3$G$  _4$   &  3.53686 &  3.53711 &  3.52297    &  7.037$-$10 \\
  108  &    2p4f    &  $^3$G$  _5$   &  3.53686 &  3.53754 &  3.52341    &  7.016$-$10 \\
  109  &    2p4f    &  $^1$G$  _4$   &          &  3.54008 &  3.52645    &  8.822$-$10 \\
  110  &    2p4f    &  $^3$D$  _3$   &  3.55092 &  3.54326 &  3.52920    &  6.765$-$10 \\
  111  &    2p4f    &  $^3$D$  _2$   &  3.55140 &  3.54355 &  3.52950    &  6.764$-$10 \\
  112  &    2p4f    &  $^3$D$  _1$   &  3.55153 &  3.54378 &  3.52973    &  6.761$-$10 \\
  113  &    2p4f    &  $^1$D$  _2$   &          &  3.54683 &  3.53305    &  7.084$-$10 \\
  114  &    2p4d    &  $^1$F$^o_3$   &          &  3.56068 &  3.54287    &  1.394$-$10 \\
  115  &    2p4d    &  $^1$P$^o_1$   &          &  3.56868 &  3.55002    &  2.266$-$10 \\
  116  &    2p4p    &  $^1$S$  _0$   &          &  3.57681 &  3.55737    &  1.482$-$09 \\
  117  &    2p5s    &  $^3$P$^o_0$   &          &  3.68170 &  3.66788    &  7.601$-$10 \\
  118  &    2p5s    &  $^3$P$^o_1$   &          &  3.68200 &  3.66819    &  7.576$-$10 \\
  119  &    2p5s    &  $^3$P$^o_2$   &          &  3.68263 &  3.66886    &  7.550$-$10 \\
  120  &    2p5s    &  $^1$P$^o_1$   &          &  3.69465 &  3.68074    &  3.997$-$10 \\
  121  &    2p5p    &  $^1$P$  _1$   &  3.71278 &  3.70468 &  3.69117    &  2.670$-$10 \\
  122  &    2p5p    &  $^3$D$  _1$   &  3.71638 &  3.70779 &  3.69433    &  7.797$-$10 \\
  123  &    2p5p    &  $^3$D$  _2$   &  3.71638 &  3.70808 &  3.69463    &  7.938$-$10 \\
  124  &    2p5p    &  $^3$D$  _3$   &  3.71638 &  3.70858 &  3.69515    &  7.934$-$10 \\
  125  &    2p5p    &  $^3$S$  _1$   &          &  3.71476 &  3.70116    &  3.248$-$10 \\
  126  &    2p5p    &  $^3$P$  _0$   &  3.72640 &  3.71875 &  3.70457    &  1.084$-$09 \\
  127  &    2p5p    &  $^3$P$  _1$   &  3.72640 &  3.71895 &  3.70479    &  1.066$-$09 \\
  128  &    2p5p    &  $^3$P$  _2$   &  3.72640 &  3.71923 &  3.70510    &  1.083$-$09 \\
  129  &    2p5d    &  $^1$D$^o_2$   &  3.73331 &  3.72472 &  3.71123    &  7.188$-$10 \\
  130  &    2p5d    &  $^3$F$^o_2$   &          &  3.72544 &  3.71206    &  1.237$-$09 \\
  131  &    2p5d    &  $^3$F$^o_3$   &          &  3.72558 &  3.71223    &  1.786$-$09 \\
  132  &    2p5d    &  $^3$F$^o_4$   &          &  3.72602 &  3.71270    &  1.809$-$09 \\
  133  &    2p5p    &  $^1$D$  _2$   &  3.73169 &  3.73175 &  3.71533    &  5.258$-$10 \\
  134  &    2p5d    &  $^3$D$^o_1$   &  3.74153 &  3.73248 &  3.71838    &  2.344$-$10 \\
  135  &    2p5d    &  $^3$D$^o_2$   &  3.74153 &  3.73259 &  3.71850    &  2.350$-$10 \\
  136  &    2p5d    &  $^3$D$^o_3$   &  3.74153 &  3.73278 &  3.71870    &  2.345$-$10 \\
  137  &    2p5f    &  $^1$F$  _3$   &          &  3.73419 &  3.72026    &  1.024$-$09 \\
  138  &    2p5f    &  $^3$F$  _2$   &  3.74406 &  3.73453 &  3.72064    &  1.043$-$09 \\
  139  &    2p5f    &  $^3$F$  _3$   &  3.74406 &  3.73461 &  3.72072    &  1.048$-$09 \\
  140  &    2p5f    &  $^3$F$  _4$   &  3.74406 &  3.73469 &  3.72082    &  1.050$-$09 \\
  141  &    2p5g    &  $^3$G$^o_4$   &  3.74626 &  3.73626 &  3.72235    &  1.469$-$09 \\
  142  &    2p5g    &  $^3$G$^o_3$   &  3.74626 &  3.73626 &  3.72236    &  1.470$-$09 \\
  143  &    2p5g    &  $^1$G$^o_4$   &  3.74626 &  3.73638 &  3.72247    &  1.487$-$09 \\
  144  &    2p5g    &  $^3$G$^o_5$   &          &  3.73638 &  3.72248    &  1.488$-$09 \\
  145  &    2p5d    &  $^3$P$^o_2$   &  3.74432 &  3.73645 &  3.72224    &  3.287$-$10 \\
  146  &    2p5d    &  $^3$P$^o_1$   &  3.74432 &  3.73673 &  3.72252    &  3.284$-$10 \\
  147  &    2p5d    &  $^3$P$^o_0$   &  3.74432 &  3.73687 &  3.72267    &  3.286$-$10 \\
  148  &    2p5f    &  $^3$G$  _3$   &  3.74366 &  3.73773 &  3.72370    &  1.269$-$09 \\
  149  &    2p5f    &  $^3$G$  _4$   &  3.74366 &  3.73798 &  3.72396    &  1.275$-$09 \\
  150  &    2p5g    &  $^3$H$^o_4$   &  3.74585 &  3.73846 &  3.72461    &  1.967$-$09 \\
\hline
\end{tabular} 
\end{table*}

\setcounter{table}{0}
\begin{table*}
\caption{... continued}
\begin{tabular}{rllrrrrrl} \hline
Index  & Configuration       & Level          &  NIST      &   GRASP   &  AS     & $\tau$ (s) \\
\hline 
  151  &    2p5g    &  $^3$H$^o_5$   &  3.74585 &  3.73848 &  3.72464    &  1.974$-$09 \\
  152  &    2p5f    &  $^3$G$  _5$   &  3.74366 &  3.73839 &  3.72439    &  1.272$-$09 \\
  153  &    2p5g    &  $^3$H$^o_6$   &  3.74585 &  3.73907 &  3.72525    &  1.995$-$09 \\
  154  &    2p5g    &  $^1$H$^o_5$   &          &  3.73910 &  3.72529    &  2.004$-$09 \\
  155  &    2p5g    &  $^3$F$^o_4$   &  3.74925 &  3.73986 &  3.72604    &  1.364$-$09 \\
  156  &    2p5g    &  $^1$F$^o_3$   &          &  3.73988 &  3.72607    &  1.367$-$09 \\
  157  &    2p5f    &  $^1$G$  _4$   &          &  3.74023 &  3.72646    &  1.525$-$09 \\
  158  &    2p5g    &  $^3$F$^o_2$   &  3.74925 &  3.74036 &  3.72656    &  1.360$-$09 \\
  159  &    2p5g    &  $^3$F$^o_3$   &  3.74925 &  3.74037 &  3.72658    &  1.362$-$09 \\
  160  &    2p5f    &  $^3$D$  _3$   &          &  3.74175 &  3.72769    &  8.861$-$10 \\
  161  &    2p5f    &  $^3$D$  _2$   &          &  3.74203 &  3.72798    &  8.823$-$10 \\
  162  &    2p5f    &  $^3$D$  _1$   &          &  3.74225 &  3.72821    &  8.842$-$10 \\
  163  &    2p5f    &  $^1$D$  _2$   &          &  3.74516 &  3.73082    &  6.213$-$10 \\
  164  &    2p5d    &  $^1$F$^o_3$   &          &  3.75389 &  3.73497    &  1.342$-$10 \\
  165  &    2p5d    &  $^1$P$^o_1$   &          &  3.76002 &  3.73894    &  1.739$-$10 \\
  166  &    2p5p    &  $^1$S$  _0$   &          &  3.78461 &  3.74510    &  7.954$-$10 \\
\hline  
\end{tabular}

\begin{flushleft}
{\small
NIST: {\tt http://www.nist.gov/pml/data/asd.cfm} \\
GRASP: Energies from the {\sc grasp} code for 166 level calculations \\
AS: Energies from the AS for 238 level calculations \\
}
\end{flushleft}
\end{table*}

\begin{figure*}
\includegraphics[angle=-90,width=0.9\textwidth]{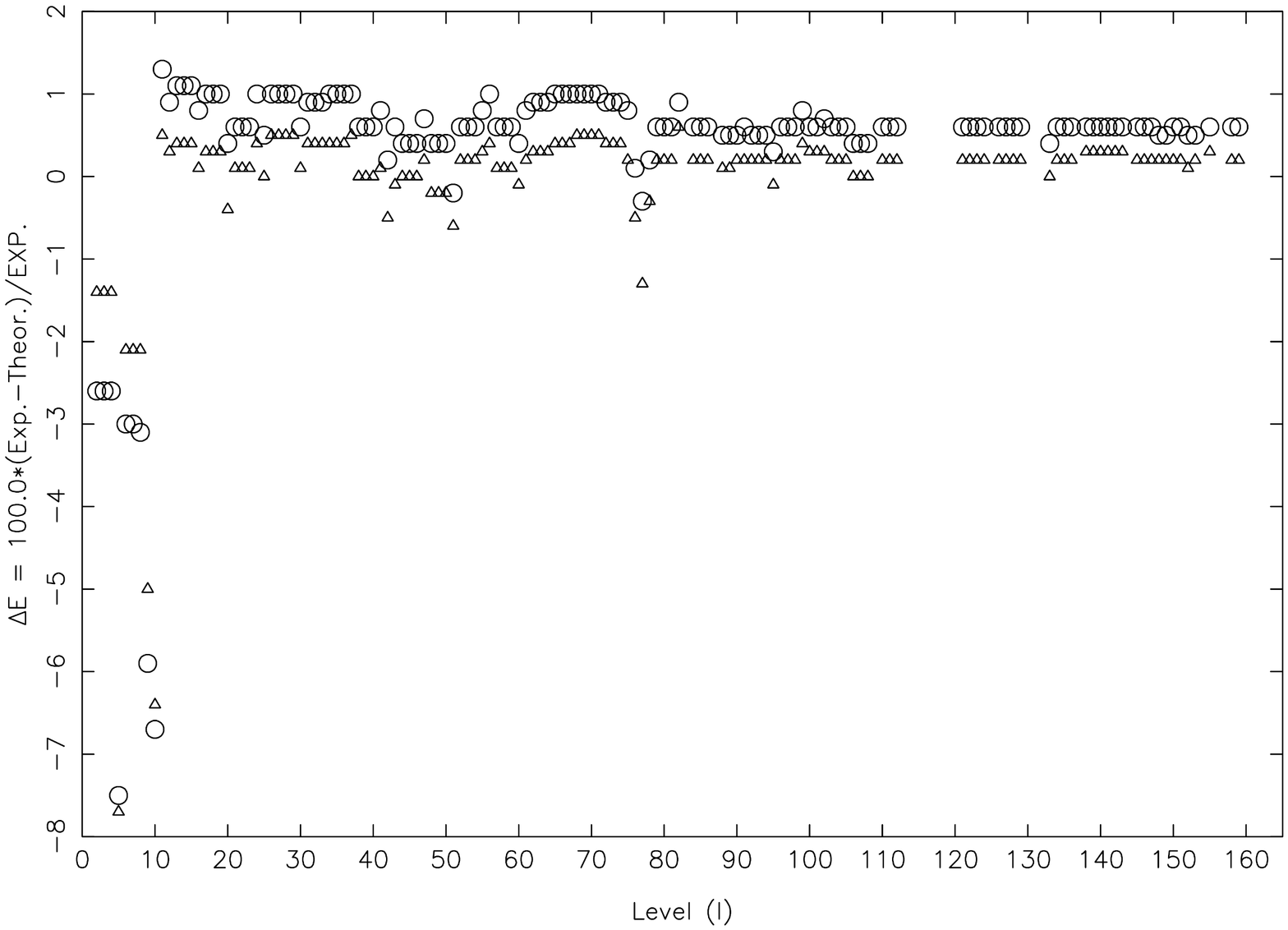}
 \vspace{-1.5cm}
 \caption{Percentage differences between experimental and theoretical energy levels. Triangles: present calculations with GRASP and circles: calculations of Fern{\'a}ndez-Menchero et al. (2014) with AS.}
\end{figure*}

\setcounter{table}{1}                                                                                                                                           
\begin{table*}                                                                                                                                                  
\caption{Transition wavelengths ($\lambda_{ij}$ in $\rm \AA$), radiative rates (A$_{ji}$ in s$^{-1}$), oscillator strengths (f$_{ij}$, dimensionless), and line  
strengths (S, in atomic units) for electric dipole (E1), and A$_{ji}$ for E2, M1 and M2 transitions in C III. ($a{\pm}b \equiv a{\times}$10$^{{\pm}b}$). See Table 1 for level indices.}     
\begin{tabular}{rrrrrrrrr}                                                                                                                                      
\hline                                                                                                                                                                                                                                                                                                               
$i$ & $j$ & $\lambda_{ij}$ & A$^{{\rm E1}}_{ji}$  & f$^{{\rm E1}}_{ij}$ & S$^{{\rm E1}}$ & A$^{{\rm E2}}_{ji}$  & A$^{{\rm M1}}_{ji}$ & A$^{{\rm M2}}_{ji}$ \\  
\hline                                                                                                                                                   
    1 &    3 &  1.882$+$03 &  6.338$+$01 &  1.010$-$07 &  6.260$-$07 &  0.000$+$00 &  0.000$+$00 &  0.000$+$00 \\       
    1 &    4 &  1.881$+$03 &  0.000$+$00 &  0.000$+$00 &  0.000$+$00 &  0.000$+$00 &  0.000$+$00 &  5.390$-$03 \\       
    1 &    5 &  9.073$+$02 &  2.152$+$09 &  7.966$-$01 &  2.379$+$00 &  0.000$+$00 &  0.000$+$00 &  0.000$+$00 \\       
    1 &    7 &  7.128$+$02 &  0.000$+$00 &  0.000$+$00 &  0.000$+$00 &  0.000$+$00 &  8.478$-$04 &  0.000$+$00 \\       
    1 &    8 &  7.126$+$02 &  0.000$+$00 &  0.000$+$00 &  0.000$+$00 &  1.953$-$02 &  0.000$+$00 &  0.000$+$00 \\       
    1 &    9 &  6.530$+$02 &  0.000$+$00 &  0.000$+$00 &  0.000$+$00 &  3.605$+$03 &  0.000$+$00 &  0.000$+$00 \\       
    ... & \\
    ... & \\
    ... & \\
\hline                                                                                                                                                          
\end{tabular}                                                                                                                                                   
\end{table*}                                           

\section[]{Energy levels}

Our calculations for C~III are larger than those performed by us for other Be-like ions. For this ion we have considered the energetically  lowest 166 levels  belonging to the 27  configurations: (1s$^2$) 2$\ell$2$\ell'$, 2$\ell$3$\ell'$,  2$\ell$4$\ell'$  and 2$\ell$5$\ell'$, while for  others (see \cite{belike} and references therein) the 68 levels of the 2$\ell$5$\ell'$  configurations were not considered. Our level energies calculated from {\sc grasp},   with the inclusion of Breit and QED (quantum electrodynamic) effects, are listed in Table~1 along with those of NIST and \cite{icft}.  

Experimental energies are available for most of the levels, with a few exceptions  such as 113--120. In Fig.~1 we show the differences (in percentage terms) between the experimental and theoretical energy levels. There is general agreement (within 1\% for all levels) between our  GRASP and the earlier AS  energies of \cite{icft}, although our results are slightly closer to those of NIST.  Differences between the theoretical and  NIST energies are smaller than 1\% for most  levels. Unfortunately, the differences for the lowest 10 levels of the 2s2p and 2p$^2$ configurations are significantly larger, up to 8\%. Clearly there is scope for improvement which can (perhaps) be  achieved by the inclusion of {\em pseudo} orbitals, as  undertaken by \cite{kab2} and \cite{mit}. This approach improves the accuracy of energy levels, although discrepancies with measurements remain up to 6\% for a few, see for example the 2p$^2$ $^1$S$_0$ level in table~1 of \cite{mit}. Therefore, there is little advantage in using pseudo orbitals, particularly because their inclusion gives rise to pseudo resonances in the subsequent scattering calculations for $\Omega$. For this reason, neither ourselves nor \cite{icft} have included  pseudo orbitals in the generation of wavefunctions. 

\section{Radiative rates}

We have calculated  A-values  for four types of transitions, namely electric dipole (E1), electric quadrupole (E2), magnetic dipole (M1) and  magnetic quadrupole (M2) as all may be required in a plasma model (see e.g. \cite{tixix} and references therein). The absorption oscillator strength ($f_{ij}$) and radiative rate A$_{ji}$ (in s$^{-1}$) for all types of  transition $i \to j$ are related by the following expression:

\begin{equation}
f_{ij} = \frac{mc}{8{\pi}^2{e^2}}{\lambda^2_{ji}} \frac{{\omega}_j}{{\omega}_i} A_{ji}
 = 1.49 \times 10^{-16} \lambda^2_{ji}  \frac{{\omega}_j}{{\omega}_i}  A_{ji} 
\end{equation}
where $m$ and $e$ are the electron mass and charge, respectively, $c$ the velocity of light,  $\lambda_{ji}$  the transition energy/wavelength in $\rm \AA$, and $\omega_i$ and $\omega_j$  the statistical weights of the lower ($i$) and upper ($j$) levels, respectively. However, the relationships between oscillator strength f$_{ij}$ (dimensionless) and the line strength S (in atomic unit, equivalent to 6.460$\times$10$^{-36}$ cm$^2$ esu$^2$) with the A-values  differ for different types of transitions -- see Eqs. (2--5) of  \cite{tixix}.

Our calculated energies/wavelengths ($\lambda$, in $\rm \AA$), radiative rates (A$_{ji}$, in s$^{-1}$), oscillator strengths (f$_{ij}$, dimensionless), and line strengths (S, in atomic unit) are listed in Table~2 for all 3802 E1 transitions among the 166 levels of  C~III.  The {\em indices} used  to represent the lower and upper levels of a transition are defined in Table~1. Similarly, there  are 4909 E2, 3728 M1  and 4975 M2 transitions  among the same 166 levels. However, for these only the A-values are listed in Table~2. Corresponding data for f- values can be easily obtained through Eq. (1). Furthermore,  results are only listed in the length  form, which are considered to be comparatively more accurate. Nevertheless, below we discuss  the velocity/length  form ratio (R), as this provides some assessment of  the accuracy of the data. 

Ideally, R should be close to unity but often is not in (almost all) large calculations, such as the present one. As for other Be-like ions \citep{belike}, C~III is no exception and hence the conclusions are similar. Specifically,  for 453 strong E1 transitions (f $\ge$ 0.01) R lies outside the range 0.8--1.2, i.e. is  more than 20\% away from unity. For most such transitions  R is within a factor of two,  but there are exceptions for a few, such as 5--121 (f = 0.05), 9--61 (f =  0.05), 19--145 (f = 0.01), 20--165 (f = 0.02) and 25--75 (f = 0.01).  For these  transitions unfortunately R is up to an order of magnitude. Similarly, for a few weaker transitions the two forms of f- values differ by up to several orders of magnitude, but most of these have very small f-values ($\sim$10$^{-5}$ or less). Nevertheless, such weak transitions may occasionally be important for the calculations of lifetimes, but should not significantly affect the modelling of plasmas.

Most of the A-values available in the literature involve levels of the $n \le$ 3 configurations. However, \cite{icft} have reported  results for a larger number of E1 transitions. For most  there is satisfactory  agreement between the two calculations, but for a few weak(er) ones  there are discrepancies  of over 50\%, as shown  in Table~3. Such discrepancies between any two calculations often arise for weak(er) transitions, mainly due to the different levels of CI included and/or the method adopted. The A-values for a few M1 transitions are also available in the literature, mainly by \cite{glass} and \cite{uis}. In Table~4 we compare our A-values for the common transitions. As expected, all such transitions are weak and there is generally no large discrepancy.  However, for the 3--5 transition there is a significant  differences  in the  A-values, as our result  is larger than of  \cite{uis}  by a factor of two, but is lower than of \cite{glass}  by a factor of four. However, \cite{ns} calculated an A-value of 1.09$\times$10$^{-3}$ s$^{-1}$ for this transition, which agrees within 20\% with our result.

\setcounter{table}{2}
\begin{table*}
\caption{Comparison of A-values for a few E1 transitions of C III.  ($a{\pm}b \equiv a{\times}$10$^{{\pm}b}$).   See Table 1 for level indices.}
\begin{tabular}{rrrrrrl} \hline
I & J& f (GRASP) & A (GRASP) & A (AS) & R  \\
\hline
    1  &    3  &  1.010$-$07  &  6.338$+$01  &  1.120$+$02  &    1.8 \\
    1  &   28  &  8.452$-$09  &  1.880$+$02  &  4.870$+$02  &    2.6 \\
    1  &   61  &  7.696$-$02  &  2.009$+$09  &  1.280$+$09  &    1.6 \\
    1  &   63  &  2.386$-$06  &  6.245$+$04  &  3.300$+$04  &    1.9 \\
    1  &   76  &  2.064$-$04  &  5.573$+$06  &  7.930$+$07  &   14.2 \\
    2  &   38  &  4.312$-$04  &  6.988$+$06  &  1.610$+$07  &    2.3 \\
    3  &    9  &  3.412$-$07  &  1.366$+$03  &  2.260$+$03  &    1.7 \\
    3  &   10  &  2.872$-$08  &  1.144$+$03  &  1.750$+$03  &    1.5 \\
    3  &   12  &  4.814$-$09  &  3.604$+$02  &  2.230$+$02  &    1.6 \\
    3  &   20  &  5.632$-$08  &  1.135$+$03  &  2.390$+$03  &    2.1 \\
    3  &   26  &  3.621$-$11  &  4.789$+$00  &  3.000$+$02  &   62.6 \\
    3  &   38  &  1.009$-$04  &  4.906$+$06  &  1.150$+$07  &    2.3 \\
    3  &   39  &  3.911$-$04  &  1.141$+$07  &  2.470$+$07  &    2.2 \\
    3  &   75  &  5.931$-$08  &  2.035$+$03  &  8.050$+$02  &    2.5 \\
    3  &   77  &  1.780$-$08  &  3.132$+$03  &  5.880$+$03  &    1.9 \\
    4  &    9  &  2.272$-$06  &  1.514$+$04  &  2.310$+$04  &    1.5 \\
    4  &   20  &  9.300$-$10  &  3.122$+$01  &  1.980$+$01  &    1.6 \\
    4  &   38  &  3.531$-$06  &  2.860$+$05  &  7.040$+$05  &    2.5 \\
    4  &   39  &  6.941$-$05  &  3.373$+$06  &  7.550$+$06  &    2.2 \\
    4  &   40  &  5.495$-$04  &  1.908$+$07  &  3.850$+$07  &    2.0 \\
    4  &   68  &  4.182$-$13  &  1.695$-$02  &  4.130$-$02  &    2.4 \\
    5  &    6  &  5.620$-$08  &  1.015$+$02  &  2.120$+$02  &    2.1 \\
    5  &    7  &  6.737$-$09  &  4.064$+$00  &  8.780$+$00  &    2.2 \\
    5  &    8  &  8.681$-$07  &  3.150$+$02  &  5.670$+$02  &    1.8 \\
    5  &   11  &  4.830$-$08  &  5.178$+$02  &  9.030$+$02  &    1.7 \\
    5  &   17  &  6.667$-$08  &  1.122$+$03  &  1.750$+$03  &    1.6 \\
    5  &   18  &  6.090$-$08  &  6.152$+$02  &  1.690$+$03  &    2.7 \\
    5  &   24  &  8.707$-$09  &  2.274$+$02  &  4.170$+$00  &   54.5 \\
    5  &   32  &  1.532$-$07  &  2.706$+$03  &  1.190$+$03  &    2.3 \\
    5  &   46  &  1.813$-$06  &  3.495$+$04  &  5.660$+$04  &    1.6 \\
    5  &   56  &  1.935$-$09  &  6.722$+$01  &  2.260$+$01  &    3.0 \\
    5  &   65  &  4.203$-$09  &  1.531$+$02  &  3.870$+$02  &    2.5 \\
    5  &   66  &  5.046$-$10  &  1.103$+$01  &  2.270$+$02  &   20.6 \\
    5  &   75  &  9.435$-$02  &  2.097$+$09  &  1.160$+$09  &    1.8 \\
    5  &   77  &  8.216$-$03  &  9.429$+$08  &  2.600$+$08  &    3.6 \\
\hline  
\end{tabular}

\begin{flushleft}
{\small
GRASP: Present calculations with the {\sc grasp} code \\
AS: Calculations of \cite{icft} with the {\sc as} code \\
R: ratio of  GRASP and AS A-values, the larger of the two is always in the nemerator \\
}
\end{flushleft}
\end{table*}

\setcounter{table}{3}
\begin{table*}
\caption{Comparison of A-values for a few M1 transitions of C~III.  ($a{\pm}b \equiv a{\times}$10$^{{\pm}b}$).  See Table 1 for level indices.}
\begin{tabular}{rrccc} \hline
I & J&  Present &  \cite{glass}  &   \cite{uis}  \\
\hline
1  &   7  &   8.478$-$4   &   5.335$-$4 &    6.53$-$4\\
2  &   3  &   1.697$-$7   &   2.162$-$7 &    2.68$-$7\\
2  &   5  &   1.480$-$3   &   1.592$-$3 &    6.12$-$3\\
3  &   4  &   1.948$-$6   &   2.139$-$6 &    2.50$-$6\\
3  &   5  &   1.293$-$3   &   5.676$-$3 &    5.52$-$4\\
4  &   5  &   1.943$-$3   &   1.981$-$3 &    8.28$-$4\\
6  &   7  &   3.483$-$7   &   4.092$-$7 &    4.52$-$7\\
7  &   8  &   9.745$-$7   &   1.316$-$6 &    1.73$-$6\\
7  &   9  &   2.455$-$4   &   1.982$-$4 &    1.16$-$5\\
7  &  10  &   1.483$-$2   &   1.924$-$2 &    4.68$-$3\\
8  &   9  &   7.578$-$4   &   6.296$-$4 &    3.64$-$5\\
\hline  
\end{tabular}

\begin{flushleft}
{\small

}
\end{flushleft}
\end{table*}

\setcounter{table}{4}
\begin{table*}
\caption{Comparison of lifetimes ($\tau$, ns) for a few levels of C~III. } 
\begin{tabular}{lrllrrl} \hline
Configuration/Level  &  Present  & \cite{nl}  \\
\hline
 2s2p $^1$P$^o$   & 0.465   &   0.50--0.66   \\
 2p$^2$ $^3$P     & 0.700   &   0.74--0.90   \\
 2p$^2$ $^1$D     & 7.464   &   6.90--9.30   \\
 2p$^2$ $^1$S     & 0.372   &   0.39--0.58   \\
 2s3s $^3$S       & 0.264   &   0.32--0.37   \\
 2s3s $^1$S       & 1.099   &   0.61--1.25   \\
 2s3p $^1$P$^o$   & 0.266   &   0.255--0.370 \\
 2s3p $^3$P$^o$   & 13.100   &   12.5--15.1   \\
 2s3d $^3$D       & 0.096   &   0.115--0.120 \\
 2s3d $^1$D       & 0.155   &   0.14--0.30   \\
 2p3s $^3$P$^o$   & 0.322   &   0.374--0.400 \\
 2p3s $^1$P$^o$   & 0.241   &   0.26--0.93   \\
 2p3p $^1$P       & 0.297   &   0.59         \\
 2p3p $^3$S       & 0.498   &   2.3          \\
 2p3p $^3$P       & 0.582   &   0.45--0.70   \\
 2p3p $^1$D       & 0.440   &   0.38--0.53   \\
 2p3p $^1$S       & 0.836   &   1.3          \\
 2p3d $^3$F$^o$   & 1.921   &   2.3          \\
 2p3d $^3$D$^o$   & 0.080   &   0.088--0.100 \\
 2p3d $^1$D$^o$   & 0.189   &   0.140--0.191 \\
 2p3d $^3$P$^o$   & 0.224   &   0.179--2.500 \\
 2p3d $^1$F$^o$   & 0.131   &   0.120--0.124 \\
 2p3d $^1$P$^o$   & 0.155   &   7.8          \\
 2s4s $^3$S       & 0.379   &   0.45--0.56   \\
 2s4p $^1$P$^o$   & 0.621   &   0.56--0.60   \\
 2s4d $^3$D       & 0.184   &   0.175--2.900 \\
 2s4d $^1$D       & 0.239   &   0.35--0.36   \\
 2s4f $^3$F$^o$   & 1.218   &   1.5--1.6     \\
 2s4f $^1$F$^o$   & 0.557   &   0.32--1.00   \\
 2s5s $^3$S       & 1.150   &   1.3          \\
 2s5p $^3$P$^o$   & 0.390   &   0.33--0.90   \\
 2s5d $^3$D       & 0.391   &   0.50         \\
 2s5d $^1$D       & 0.391   &   0.82         \\
 2s5f $^3$F$^o$   & 1.077   &   1.0-1.2      \\
 2s5f $^1$F$^o$   & 0.244   &   0.3          \\
 2s5g $^3$G       & 2.899   &   3.3          \\
 2s5g $^1$G       & 2.899   &   3.6          \\
 2p5g $^3$G$^o$   & 1.470   &   0.7          \\
\hline  
\end{tabular}

\begin{flushleft}
{\small

}
\end{flushleft}
\end{table*}

\setcounter{table}{5}
\begin{table*}
\caption{Comparison of lifetimes ($\tau$, s)  for the lowest 20 levels of C~III.  ($a{\pm}b \equiv a{\times}$10$^{{\pm}b}$). }
\begin{tabular}{rllrllrrl} \hline
Index  & Configuration       & Level & Present      &     \cite{tff}      \\
\hline
    1  &    2s$^2$  &  $^1$S$  _0$   &  .....       &  .....       \\
    2  &    2s2p    &  $^3$P$^o_0$   &  .....       &  .....       \\
    3  &    2s2p    &  $^3$P$^o_1$   &  1.578$-$02  &  9.616$-$03  \\
    4  &    2s2p    &  $^3$P$^o_2$   &  1.854$+$02  &  1.916$+$02  \\
    5  &    2s2p    &  $^1$P$^o_1$   &  4.648$-$10  &  5.615$-$10  \\
    6  &    2p$^2$  &  $^3$P$  _0$   &  6.993$-$10  &  7.571$-$10  \\
    7  &    2p$^2$  &  $^3$P$  _1$   &  6.990$-$10  &  7.567$-$10  \\
    8  &    2p$^2$  &  $^3$P$  _2$   &  6.985$-$10  &  7.562$-$10  \\
    9  &    2p$^2$  &  $^1$D$  _2$   &  7.464$-$09  &  7.191$-$09  \\
   10  &    2p$^2$  &  $^1$S$  _0$   &  3.718$-$10  &  4.764$-$10  \\ 
   11  &    2s3s    &  $^3$S$  _1$   &  2.635$-$10  &  2.709$-$10  \\
   12  &    2s3s    &  $^1$S$  _0$   &  1.099$-$09  &  1.171$-$09  \\
   13  &    2s3p    &  $^3$P$^o_0$   &  1.312$-$08  &  1.352$-$08  \\
   14  &    2s3p    &  $^3$P$^o_1$   &  4.252$-$09  &  1.340$-$08  \\
   15  &    2s3p    &  $^3$P$^o_2$   &  1.309$-$08  &  1.347$-$08  \\
   16  &    2s3p    &  $^1$P$^o_1$   &  2.657$-$10  &  2.512$-$10  \\
   17  &    2s3d    &  $^3$D$  _1$   &  9.623$-$11  &  9.400$-$11  \\
   18  &    2s3d    &  $^3$D$  _2$   &  9.625$-$11  &  9.402$-$11  \\
   19  &    2s3d    &  $^3$D$  _3$   &  9.628$-$11  &  9.404$-$11  \\
   20  &    2s3d    &  $^1$D$  _2$   &  1.551$-$10  &  1.589$-$10  \\
\hline  
\end{tabular}

\begin{flushleft}
{\small

}
\end{flushleft}

\end{table*}

\setcounter{table}{6}     
\begin{table*}      
\caption{Collision strengths for resonance transitions of  C III. ($a{\pm}b \equiv$ $a\times$10$^{{\pm}b}$). See Table 1 for level indices.}          
\begin{tabular}{rrlllllllr}                                                                                   
\hline                                                                                                                                                                                                             
\multicolumn{2}{c}{Transition} & \multicolumn{4}{c}{Energy (Ryd)}\\                                           
\hline                                                                                                        
  $i$ & $j$ &    5  &  10  &  15  &   20    \\                                                                
\hline 
  1 &  2 &  2.691$-$02 &  1.153$-$02 &  5.983$-$03 &  3.613$-$03 \\
  1 &  3 &  8.073$-$02 &  3.460$-$02 &  1.795$-$02 &  1.084$-$02 \\
  1 &  4 &  1.345$-$01 &  5.763$-$02 &  2.990$-$02 &  1.805$-$02 \\
  1 &  5 &  7.240$+$00 &  9.860$+$00 &  1.144$+$01 &  1.260$+$01 \\
  1 &  6 &  5.215$-$04 &  1.890$-$04 &  8.446$-$05 &  4.416$-$05 \\
  1 &  7 &  1.563$-$03 &  5.661$-$04 &  2.524$-$04 &  1.317$-$04 \\
  1 &  8 &  2.607$-$03 &  9.456$-$04 &  4.232$-$04 &  2.221$-$04 \\
  1 &  9 &  2.938$-$01 &  2.760$-$01 &  2.646$-$01 &  2.585$-$01 \\
  1 & 10 &  7.496$-$02 &  7.864$-$02 &  7.479$-$02 &  7.049$-$02 \\
    ... & \\
    ... & \\
    ... & \\
\hline                                                                                                        
\end{tabular}                                                                                                 
\end{table*}                                                   

\section{Lifetimes}

The lifetime $\tau$ for a level $j$ is defined as follows:

\begin{equation}  {\tau}_j = \frac{1}{{\sum_{i}^{}} A_{ji}}.  
\end{equation} 
Since this is a measurable quantity, it facilitates an assessment of  the accuracy of the A-values, particularly when a single transition dominates the contributions. Therefore, in Table~1 we have also listed our calculated lifetimes. Generally, A-values for E1 transitions are considerably larger in magnitude and hence dominate  the determination of $\tau$, but for higher accuracy we have also  included the contributions from E2, M1 and M2.  Their  inclusion is particularly important for those levels for which either E1 transitions do not exist or are dominated by others.

There have been several measurements of $\tau$ for a few levels of C~III, and results up to 1990 have been compiled by \cite{nl}. In Table~5 we compare our data with their compilation. For some levels there are several measurements and therefore we have listed the {\em range} for convenience (specific results are given in  table IIIa of \cite{nl}). Most levels show reasonable agreement (within a few percent) between theory and measurement. However, there are also striking discrepancies for five, namely 2p3p $^1$P, 2p3p $^3$S, 2p3d $^1$P$^o$, 2s5f $^1$D and 2p5g~$^3$G$^o$, where  the only available measurements are from \cite{pou1} and \cite*{pou2}. For the 2p3d $^1$P$^o$ level the discrepancy is the largest (by a factor of 50), but our result of 0.155 ns agrees closely with the calculations of \cite{nl}, i.e. 0.17 ns. Similarly, for other levels where there are large discrepancies  in Table~5 there is satisfactory agreement between our results and the calculations of \cite{nl}. Therefore, for the 2p3d $^1$P$^o$ level the measured $\tau$ of \cite{pou2} appears to be incorrect.  \cite{tn} have measured $\tau$ for  2p$^2$~$^1$D and 2s3p~$^3$P$^o$ to be 6.9$\pm$0.2 and 11.6$\pm$0.3~ns, respectively and for both there is no large discrepancy with our calculations (7.5 and 13.1 ns, respectively) or other earlier experimental values. 

 \cite{tff} have calculated $\tau$ for the lowest 20 levels of C~III  and their results are compared with ours in  Table~6. The only level for which there is a serious discrepancy is 2s3p $^3$P$^o_1$, with our value lower by about a factor of three. This is because in our calculations the A(f)-value for the 2s$^2$ $^1$S$_0$--3s3p $^3$P$^o_1$ transition is 1.362$\times$10$^8$ s$^{-1}$ (9.16$\times$10$^{-3}$) while theirs is 5.18$\times$10$^5$ s$^{-1}$ (3.45$\times$10$^{-5}$). However, being intercombination it is  a weak transition and therefore its A(f) values fluctuate with differing amount of CI and methods. For example, the A-value reported by \cite{uiss} is 8.41$\times$10$^4$ s$^{-1}$ whereas \cite{cc}  calculated A = 1.0$\times$10$^2$ s$^{-1}$. Therefore, due to the paucity of measurements for this level, it is  difficult to fully assess the accuracy of any calculation. Nevertheless, since there is a general agreement between theory and measurement for most of the levels/states as shown in Tables~5 and 6, we are confident that our calculations for A, f and $\tau$ are accurate to better than 20\% for a majority of transitions (levels). 
 
 \begin{figure*}
\includegraphics[angle=90,width=0.9\textwidth]{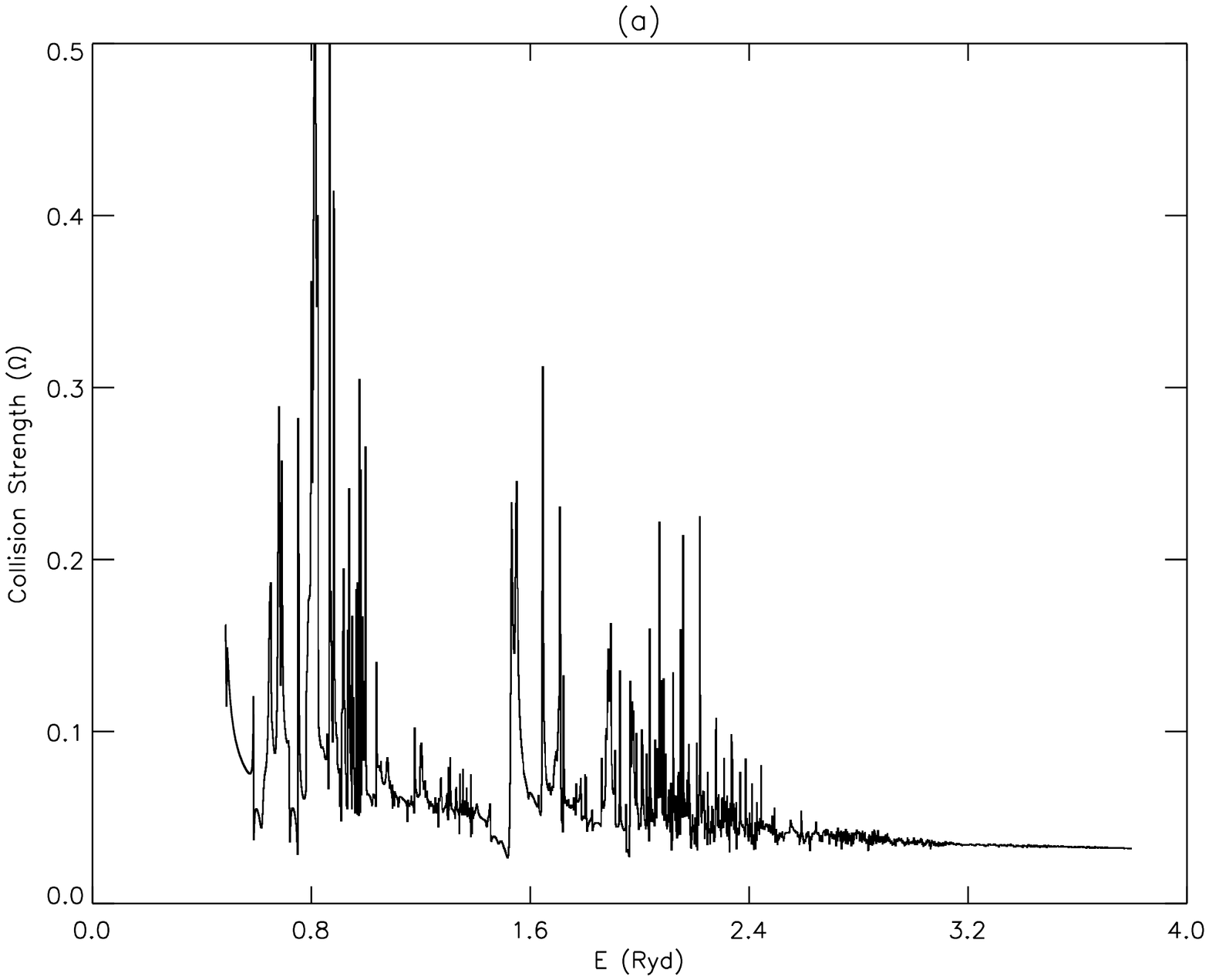}
 \vspace{-1.5cm}
 \end{figure*}

 \begin{figure*}
\includegraphics[angle=90,width=0.9\textwidth]{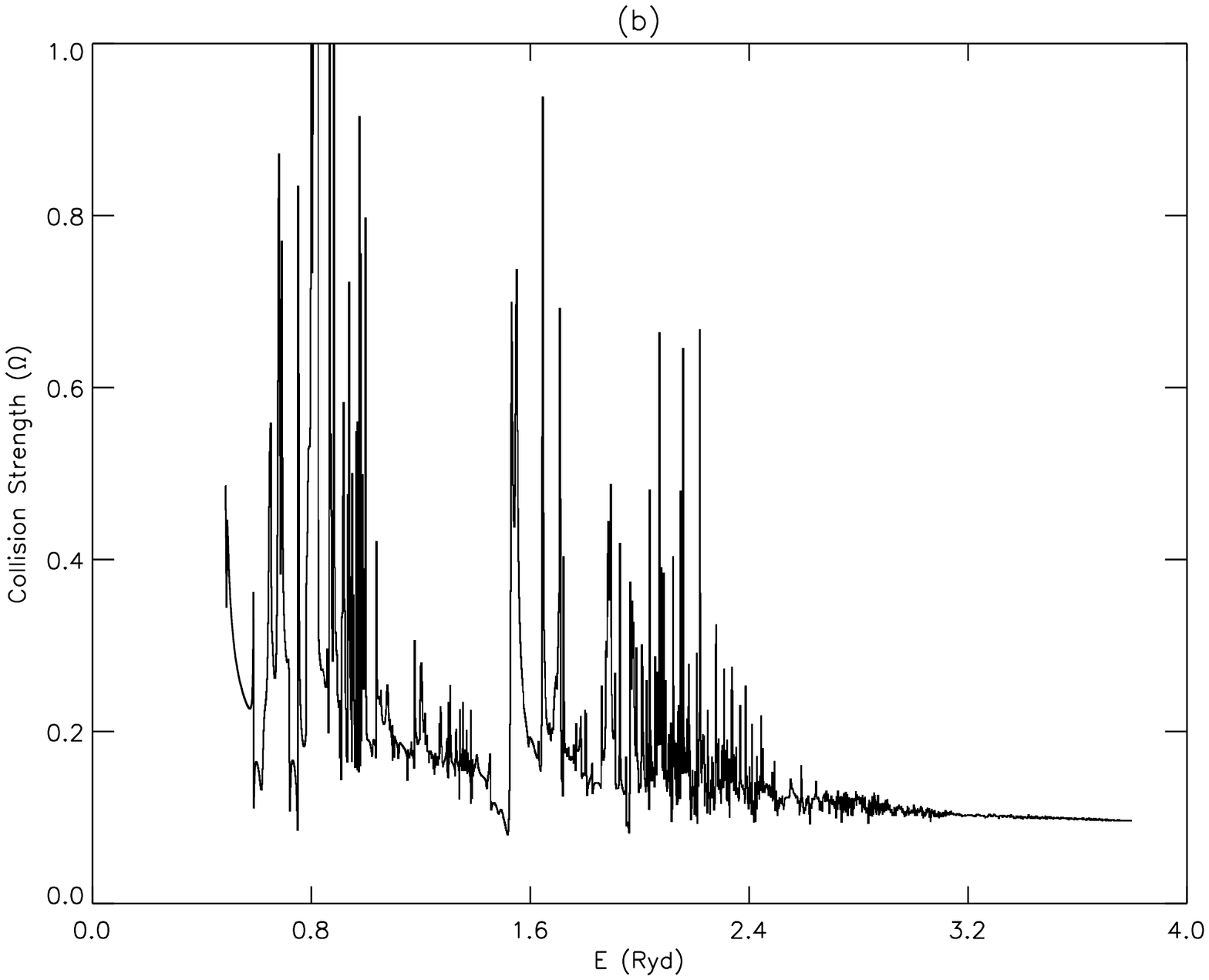}
 \vspace{-1.5cm}
 \caption{Collision strengths for the (a) 1--2 (2s$^2$ $^1$S$_0$ -- 2s2p $^3$P$^o_{0}$) and (b) 1--3 (2s$^2$ $^1$S$_0$ -- 2s2p $^3$P$^o_{1}$) transition of C III.}
 \end{figure*}

 \section{Collision strengths}

The collision strength for electron impact excitation  ($\Omega$) is  a symmetric and dimensionless quantity,  related to the better-known parameter collision cross section ($\sigma_{ij}$) -- see eq. (7) of \cite{tixix}. As stated earlier and in our work on other Be-like ions,   we have adopted the  relativistic  {\sc darc} code for calculating  $\Omega$.  It is based on the $jj$ coupling scheme and uses the  Dirac-Coulomb Hamiltonian in an $R$-matrix approach. The $R$-matrix radius adopted for  C~III is 28.0 atomic unit, and 55  continuum orbitals have been included for each channel angular momentum in the expansion of the wavefunction. This large expansion has become necessary  to compute $\Omega$ up to an energy of  21 Ryd, so that the subsequent values of effective collision strength $\Upsilon$ (see section 6)  can be reliably calculated up to T$_e$ = 8.0$\times$10$^{5}$ K, well above  the temperature of maximum abundance in ionisation equilibrium for C~III, i.e. 7.9 $\times$10$^{4}$ K  \citep*{pb}. However, considering the number of levels involved (166) the maximum number of channels generated for a partial wave is 828, and the corresponding size of the Hamiltonian (H) matrix becomes 45~714. Therefore, the present calculations are computationally more demanding and challenging than for other Be-like ions, such as Ti~XIX \citep{tixix} for which the size of H was only 23~579. Furthermore, to achieve  convergence of  $\Omega$ for a majority of transitions and at all energies, we have included all partial waves with angular momentum $J \le$ 40.5.  To account for higher neglected partial waves we have included a top-up in the same way as for other ions, i.e. the Coulomb-Bethe  \citep*{ab} and  geometric series  approximations for allowed and forbidden transitions, respectively. For most transitions and at most energies these  contributions are small but nevertheless make our calculated values of $\Omega$ comparatively more accurate.

No theoretical or experimental data for $\Omega$ are available for comparison  with  our results. However, in Table~7 we list our values of $\Omega$ for resonance transitions of  C~III at four energies {\em above} thresholds, i.e. 5, 10, 15 and 20 Ryd. The  indices used  to represent the levels of a transition have already been defined in Table~1. We hope  our results in the table for $\Omega$ will be useful for future comparison with experimental and other theoretical data. In the  threshold energy region, $\Omega$ does not  vary smoothly,  especially for (semi) forbidden transitions. There often are  numerous closed-channel (Feshbach) resonances in this region, as shown in fig.~2 of  \cite{icft} for four transitions, namely 1--3 (2s$^2$ $^1$S$_0$ -- 2s2p $^3$P$^o_{1}$), 1--4 (2s$^2$ $^1$S$_0$ -- 2s2p $^3$P$^o_{2}$), 1--5 (2s$^2$ $^1$S$_0$ -- 2s2p $^1$P$^o_{1}$) and 1--9 (2s$^2$ $^1$S$_0$ -- 2p$^2$ $^1$D$_{2}$). We observe similar resonances, in both frequency and magnitude,  and for illustration show these in Fig.~2 (a and b) for  two transitions, i.e. 1--2 (2s$^2$ $^1$S$_0$ -- 2s2p $^3$P$^o_{0}$) and 1--3 (2s$^2$ $^1$S$_0$ -- 2s2p $^3$P$^o_{1}$ ). The first  is a forbidden transition whereas the second is  inter-combination.

 \begin{figure*}
\includegraphics[angle=-90,width=0.9\textwidth]{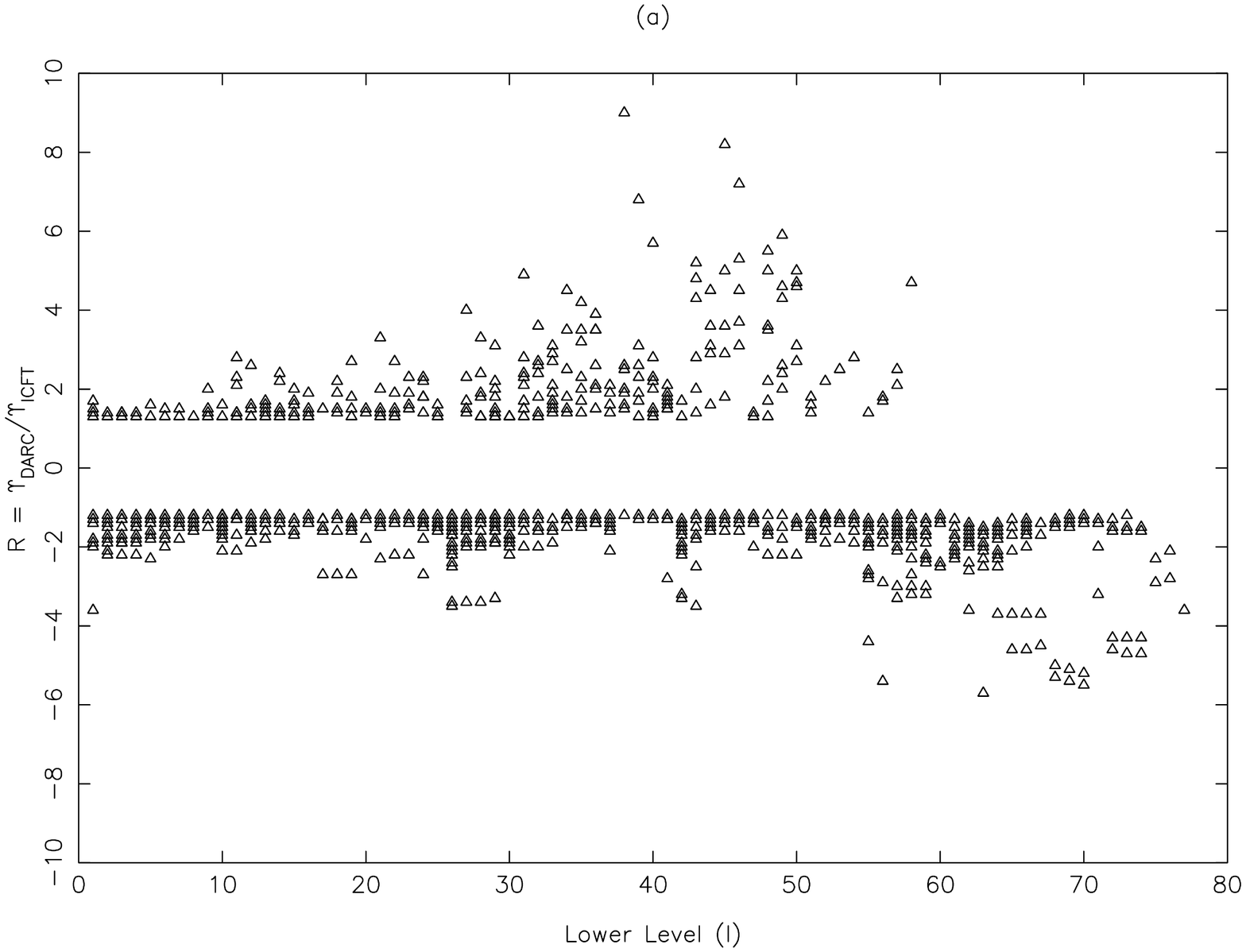}
 \vspace{-1.5cm}
 \end{figure*}

 \begin{figure*}
\includegraphics[angle=-90,width=0.9\textwidth]{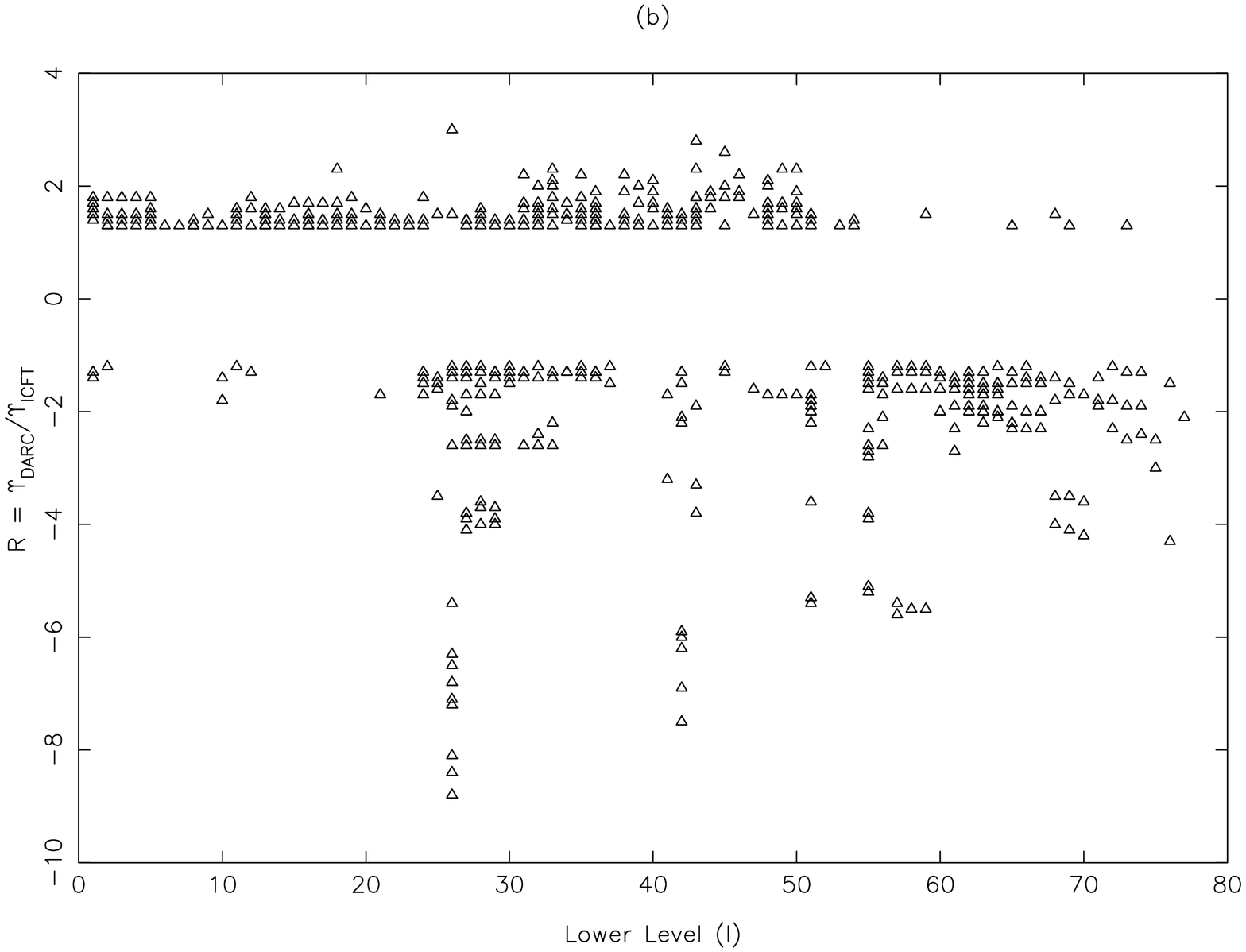}
 \vspace{-1.5cm}
 \end{figure*}

 \begin{figure*}
\includegraphics[angle=-90,width=0.9\textwidth]{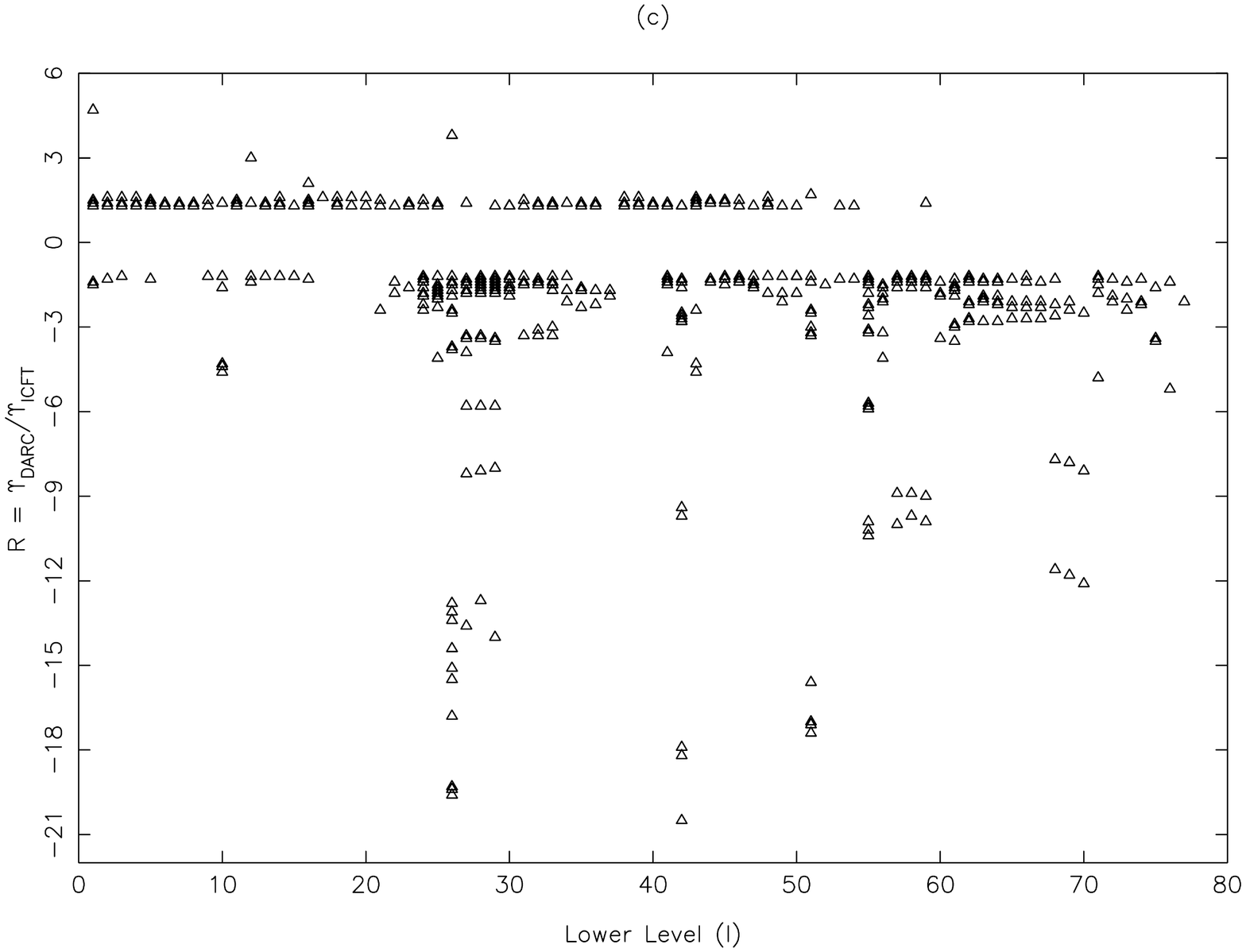}
 \vspace{-1.5cm}
 \caption{Comparison of DARC and ICFT $\Upsilon$ for transitions of C~III at (a) T$_e$ = 1800 K, (b) T$_e$ = 90~000 K and (c) T$_e$ = 450~000 K. Negative R values indicate that $\Upsilon_{\rm ICFT}$ $>$ $\Upsilon_{\rm DARC}$.}
 \end{figure*}

\section{Effective collision strengths}

Since $\Omega$ does not vary smoothly with energy, as shown in Fig. 2,  it is the {\em effective} collision strength ($\Upsilon$)  which is required in plasma modelling applications. This is determined from the collision strengths ($\Omega$) by averaging  over a suitable distribution of electron velocities.  The most appropriate and widely used distribution for applications in astrophysics is {\em Maxwellian} and hence is adopted in both  earlier and the present work, and also by \cite{icft}.  The presence  of resonances is generally significant  for forbidden, semi-forbidden and inter-combination transitions (see Fig.~2) and therefore for these the enhancement in the values of $\Upsilon$ is substantial. Similarly, values of $\Upsilon$ are affected more towards the lower range of temperatures. Such enhancements have  already been noted for a wide range of ions, including Be-like \citep{belike}. However, to account for their contribution resonances need to be  resolved in a fine energy mesh. Fortunately, resonances for  transitions in C~III are not very dense, as may be seen from our Fig.~2 and fig.~2 of \cite{icft}. Nevertheless, we have performed our calculations of $\Omega$ at  $\sim$2500 energies in the thresholds region with a mesh of (generally)  0.001 Ryd to calculate $\Upsilon$, which is given by: 

\begin{equation}
\Upsilon(T_e) = \int_{0}^{\infty} {\Omega}(E) \, {\rm exp}(-E_j/kT_e) \,d(E_j/{kT_e}),
\end{equation}
where $k$ is Boltzmann constant, T$_e$  the electron temperature in K, and E$_j$  the electron energy with respect to the final (excited) state. Once the value of $\Upsilon$ is
known the corresponding results for the excitation q(i,j) and de-excitation q(j,i) rates can be easily obtained from the following equations:

\begin{equation}
q(i,j) = \frac{8.63 \times 10^{-6}}{{\omega_i}{T_e^{1/2}}} \Upsilon \, {\rm exp}(-E_{ij}/{kT_e}) \hspace*{1.0 cm}{\rm cm^3s^{-1}}
\end{equation}
and
\begin{equation}
q(j,i) = \frac{8.63 \times 10^{-6}}{{\omega_j}{T_e^{1/2}}} \Upsilon \hspace*{1.0 cm}{\rm cm^3 s^{-1}},
\end{equation}
where $\omega_i$ and $\omega_j$ are the statistical weights of the initial ($i$) and final ($j$) states, respectively, and E$_{ij}$ is the transition energy.

\setcounter{table}{7}                                                                                                                       
\begin{table*}                                                                                                                              
\caption{Effective collision strengths for transitions in  C III. ($a{\pm}b \equiv a{\times}10^{{\pm}b}$). See Table 1 for level indices.}                                
\begin{tabular}{rrlllllllllr}                                                                                                               
\hline                                                                                                                                                                                                                                                                   
\multicolumn {2}{c}{Transition} & \multicolumn{08}{c}{Temperature (log T$_e$, K)}\\                                                         
\hline                                                                                                                                    
$i$ & $j$ &            4.50  &    4.70  &    4.90   &   5.10  &    5.30  &    5.50  &   5.70  &    5.90   \\                              
\hline                                                                                                                                                        
  1 &  2 &  1.042$-$01 &  1.028$-$01 &  9.681$-$02 &  8.742$-$02 &  7.608$-$02 &  6.402$-$02 &  5.216$-$02 &  4.107$-$02 \\
  1 &  3 &  3.132$-$01 &  3.090$-$01 &  2.907$-$01 &  2.625$-$01 &  2.284$-$01 &  1.922$-$01 &  1.566$-$01 &  1.233$-$01 \\
  1 &  4 &  5.238$-$01 &  5.161$-$01 &  4.853$-$01 &  4.380$-$01 &  3.810$-$01 &  3.206$-$01 &  2.612$-$01 &  2.056$-$01 \\
  1 &  5 &  3.588$+$00 &  3.742$+$00 &  3.961$+$00 &  4.274$+$00 &  4.720$+$00 &  5.337$+$00 &  6.124$+$00 &  6.901$+$00 \\
  1 &  6 &  3.144$-$03 &  3.041$-$03 &  2.852$-$03 &  2.546$-$03 &  2.152$-$03 &  1.733$-$03 &  1.341$-$03 &  1.002$-$03 \\
  1 &  7 &  9.452$-$03 &  9.132$-$03 &  8.558$-$03 &  7.633$-$03 &  6.449$-$03 &  5.193$-$03 &  4.017$-$03 &  3.001$-$03 \\
  1 &  8 &  1.583$-$02 &  1.528$-$02 &  1.432$-$02 &  1.277$-$02 &  1.079$-$02 &  8.684$-$03 &  6.717$-$03 &  5.017$-$03 \\
  1 &  9 &  3.409$-$01 &  3.476$-$01 &  3.477$-$01 &  3.413$-$01 &  3.314$-$01 &  3.205$-$01 &  3.094$-$01 &  2.945$-$01 \\
  1 & 10 &  8.416$-$02 &  8.256$-$02 &  7.976$-$02 &  7.702$-$02 &  7.524$-$02 &  7.465$-$02 &  7.467$-$02 &  7.354$-$02 \\
    ... & \\
    ... & \\
    ... & \\
\hline                                                                                                                                      
\end{tabular}                                                                                                                               
\end{table*}                                                               

\setcounter{table}{8}
\begin{table*}
\caption{Comparison of $\Upsilon$ for the resonance transitions of C~III at three temperatures.  ($a{\pm}b \equiv a{\times}$10$^{{\pm}b}$). See Table 1 for level indices.}
\begin{tabular}{rllrrrrrl} \hline
I & J & \multicolumn{3}{c}{DARC} & \multicolumn{3}{c}{ICFT} \\
 \multicolumn{2}{c}{T$_e$, K} &   1800           & 90~000        &  450~000      &   1800           & 90~000        &  450~000  \\
\hline
    1  &    2  &  1.075$-$1  &  9.453$-$2  &  5.488$-$2   &   9.000$-$2  &  8.980$-$2  &  5.430$-$2 \\
    1  &    3  &  3.309$-$1  &  2.839$-$1  &  1.648$-$1   &   2.850$-$1  &  2.700$-$1  &  1.630$-$1 \\
    1  &    4  &  5.816$-$1  &  4.738$-$1  &  2.748$-$1   &   4.580$-$1  &  4.490$-$1  &  2.720$-$1 \\
    1  &    5  &  3.036$+$0  &  4.035$+$0  &  5.929$+$0   &   3.000$+$0  &  3.920$+$0  &  5.860$+$0 \\
    1  &    6  &  5.310$-$3  &  2.780$-$3  &  1.428$-$3   &   4.020$-$3  &  2.980$-$3  &  1.500$-$3 \\
    1  &    7  &  1.636$-$2  &  8.342$-$3  &  4.280$-$3   &   1.230$-$2  &  8.940$-$3  &  4.510$-$3 \\
    1  &    8  &  2.879$-$2  &  1.395$-$2  &  7.156$-$3   &   2.000$-$2  &  1.490$-$2  &  7.520$-$3 \\
    1  &    9  &  2.689$-$1  &  3.465$-$1  &  3.121$-$1   &   2.680$-$1  &  3.190$-$1  &  2.950$-$1 \\
    1  &   10  &  5.194$-$2  &  7.897$-$2  &  7.468$-$2   &   7.070$-$2  &  7.520$-$2  &  7.290$-$2 \\
    1  &   11  &  2.298$-$1  &  1.090$-$1  &  4.244$-$2   &   2.950$-$1  &  1.160$-$1  &  4.340$-$2 \\
    1  &   12  &  3.059$-$1  &  2.616$-$1  &  3.212$-$1   &   3.050$-$1  &  2.680$-$1  &  3.220$-$1 \\
    1  &   13  &  1.241$-$2  &  1.237$-$2  &  6.548$-$3   &   1.650$-$2  &  1.220$-$2  &  6.420$-$3 \\
    1  &   14  &  4.078$-$2  &  3.968$-$2  &  2.476$-$2   &   5.290$-$2  &  3.860$-$2  &  2.320$-$2 \\
    1  &   15  &  6.322$-$2  &  6.231$-$2  &  3.285$-$2   &   8.360$-$2  &  6.130$-$2  &  3.210$-$2 \\
    1  &   16  &  1.078$-$1  &  8.499$-$2  &  1.168$-$1   &   1.280$-$1  &  8.630$-$2  &  1.200$-$1 \\
    1  &   17  &  3.186$-$2  &  2.946$-$2  &  1.907$-$2   &   3.330$-$2  &  2.890$-$2  &  1.830$-$2 \\
    1  &   18  &  5.331$-$2  &  4.916$-$2  &  3.183$-$2   &   5.590$-$2  &  4.820$-$2  &  3.050$-$2 \\
    1  &   19  &  7.499$-$2  &  6.872$-$2  &  4.449$-$2   &   7.740$-$2  &  6.750$-$2  &  4.260$-$2 \\
    1  &   20  &  1.124$-$1  &  1.434$-$1  &  2.849$-$1   &   1.260$-$1  &  1.430$-$1  &  2.820$-$1 \\
    1  &   21  &  5.516$-$3  &  1.793$-$3  &  7.020$-$4   &   6.260$-$3  &  1.700$-$3  &  6.850$-$4 \\
    1  &   22  &  1.813$-$2  &  5.444$-$3  &  2.147$-$3   &   1.940$-$2  &  5.130$-$3  &  2.070$-$3 \\
    1  &   23  &  3.029$-$2  &  9.091$-$3  &  3.562$-$3   &   3.300$-$2  &  8.530$-$3  &  3.420$-$3 \\
    1  &   24  &  4.441$-$2  &  2.159$-$2  &  1.084$-$2   &   5.510$-$2  &  2.020$-$2  &  1.020$-$2 \\
    1  &   25  &  8.458$-$2  &  4.426$-$2  &  4.441$-$2   &   1.060$-$1  &  4.360$-$2  &  4.520$-$2 \\
    1  &   26  &  4.746$-$2  &  6.652$-$2  &  8.800$-$2   &   4.780$-$2  &  5.840$-$2  &  8.310$-$2 \\
    1  &   27  &  7.843$-$3  &  4.494$-$3  &  2.636$-$3   &   8.730$-$3  &  3.990$-$3  &  2.360$-$3 \\
    1  &   28  &  2.406$-$2  &  1.348$-$2  &  7.905$-$3   &   2.630$-$2  &  1.200$-$2  &  7.120$-$3 \\
    1  &   29  &  4.236$-$2  &  2.268$-$2  &  1.322$-$2   &   4.350$-$2  &  2.000$-$2  &  1.180$-$2 \\
    1  &   30  &  1.169$-$2  &  3.011$-$3  &  2.394$-$3   &   2.050$-$2  &  2.990$-$3  &  2.310$-$3 \\
    1  &   31  &  9.267$-$3  &  8.622$-$3  &  5.807$-$3   &   9.530$-$3  &  7.490$-$3  &  5.220$-$3 \\
    1  &   32  &  1.697$-$2  &  1.457$-$2  &  9.780$-$3   &   1.610$-$2  &  1.250$-$2  &  8.770$-$3 \\
    1  &   33  &  2.243$-$2  &  2.046$-$2  &  1.378$-$2   &   2.230$-$2  &  1.770$-$2  &  1.230$-$2 \\
    1  &   34  &  1.084$-$2  &  5.740$-$3  &  3.189$-$3   &   9.910$-$3  &  5.500$-$3  &  3.070$-$3 \\
    1  &   35  &  1.552$-$2  &  8.202$-$3  &  4.530$-$3   &   1.400$-$2  &  7.710$-$3  &  4.300$-$3 \\
    1  &   36  &  1.939$-$2  &  1.026$-$2  &  5.694$-$3   &   1.770$-$2  &  9.890$-$3  &  5.510$-$3 \\
    1  &   37  &  3.734$-$2  &  2.516$-$2  &  2.736$-$2   &   3.330$-$2  &  2.460$-$2  &  2.730$-$2 \\
    1  &   38  &  5.753$-$3  &  4.493$-$3  &  2.722$-$3   &   5.940$-$3  &  3.970$-$3  &  2.350$-$3 \\
    1  &   39  &  9.892$-$3  &  7.475$-$3  &  4.510$-$3   &   1.010$-$2  &  6.560$-$3  &  3.880$-$3 \\
    1  &   40  &  1.458$-$2  &  1.023$-$2  &  6.142$-$3   &   1.400$-$2  &  9.050$-$3  &  5.310$-$3 \\
    1  &   41  &  3.707$-$2  &  5.159$-$2  &  1.045$-$1   &   3.890$-$2  &  4.410$-$2  &  9.340$-$2 \\
    1  &   42  &  2.605$-$2  &  2.665$-$2  &  3.244$-$2   &   2.390$-$2  &  2.450$-$2  &  3.000$-$2 \\
    1  &   43  &  9.060$-$3  &  3.404$-$3  &  1.350$-$3   &   1.010$-$2  &  3.130$-$3  &  1.240$-$3 \\
    1  &   44  &  4.974$-$4  &  2.978$-$4  &  1.634$-$4   &   7.050$-$4  &  2.660$-$4  &  1.530$-$4 \\
    1  &   45  &  1.511$-$3  &  8.960$-$4  &  4.920$-$4   &   2.140$-$3  &  7.970$-$4  &  4.590$-$4 \\
    1  &   46  &  2.712$-$3  &  1.513$-$3  &  8.223$-$4   &   3.600$-$3  &  1.330$-$3  &  7.670$-$4 \\
    1  &   47  &  5.374$-$3  &  3.090$-$3  &  3.552$-$3   &   7.680$-$3  &  2.920$-$3  &  3.260$-$3 \\
    1  &   48  &  6.940$-$3  &  3.717$-$3  &  1.851$-$3   &   6.710$-$3  &  3.490$-$3  &  1.760$-$3 \\
    1  &   49  &  1.076$-$2  &  5.240$-$3  &  2.600$-$3   &   9.600$-$3  &  4.890$-$3  &  2.470$-$3 \\
    1  &   50  &  1.572$-$2  &  6.772$-$3  &  3.352$-$3   &   1.240$-$2  &  6.280$-$3  &  3.170$-$3 \\
\hline
\end{tabular} 
\end{table*}

\setcounter{table}{8}
\begin{table*}
\caption{... continued}
\begin{tabular}{rllrrrrrl} \hline
I & J & \multicolumn{3}{c}{DARC} & \multicolumn{3}{c}{ICFT} \\
 \multicolumn{2}{c}{T$_e$, K} &   1800           & 90~000        &  450~000      &   1800           & 90~000        &  450~000  \\
\hline
    1  &   51  &  2.422$-$2  &  1.110$-$2  &  7.761$-$3   &   2.010$-$2  &  1.100$-$2  &  9.310$-$3 \\
    1  &   52  &  5.755$-$4  &  2.926$-$4  &  1.384$-$4   &   7.580$-$4  &  2.600$-$4  &  1.240$-$4 \\
    1  &   53  &  9.746$-$4  &  4.977$-$4  &  2.359$-$4   &   1.270$-$3  &  4.320$-$4  &  2.040$-$4 \\
    1  &   54  &  1.401$-$3  &  6.834$-$4  &  3.225$-$4   &   1.760$-$3  &  6.020$-$4  &  2.860$-$4 \\
    1  &   55  &  3.412$-$2  &  3.473$-$2  &  3.381$-$2   &   2.830$-$2  &  2.930$-$2  &  3.400$-$2 \\
    1  &   56  &  1.811$-$2  &  9.622$-$3  &  4.837$-$3   &   2.020$-$2  &  8.200$-$3  &  4.420$-$3 \\
    1  &   57  &  1.400$-$2  &  7.391$-$3  &  3.653$-$3   &   1.100$-$2  &  6.000$-$3  &  3.060$-$3 \\
    1  &   58  &  9.014$-$3  &  4.468$-$3  &  2.195$-$3   &   6.630$-$3  &  3.630$-$3  &  1.860$-$3 \\
    1  &   59  &  3.083$-$3  &  1.500$-$3  &  7.364$-$4   &   2.210$-$3  &  1.210$-$3  &  6.210$-$4 \\
    1  &   60  &  2.187$-$2  &  1.182$-$2  &  1.220$-$2   &   1.620$-$2  &  1.140$-$2  &  1.210$-$2 \\
    1  &   61  &  3.333$-$2  &  3.105$-$2  &  3.607$-$2   &   2.020$-$2  &  2.060$-$2  &  2.900$-$2 \\
    1  &   62  &  1.057$-$2  &  8.948$-$3  &  5.343$-$3   &   1.400$-$2  &  5.710$-$3  &  3.700$-$3 \\
    1  &   63  &  6.105$-$3  &  5.308$-$3  &  3.179$-$3   &   8.270$-$3  &  3.400$-$3  &  2.220$-$3 \\
    1  &   64  &  2.040$-$3  &  1.749$-$3  &  1.048$-$3   &   2.760$-$3  &  1.130$-$3  &  7.310$-$4 \\
    1  &   65  &  7.394$-$3  &  7.297$-$3  &  4.446$-$3   &   6.330$-$3  &  5.270$-$3  &  3.370$-$3 \\
    1  &   66  &  1.215$-$2  &  1.209$-$2  &  7.375$-$3   &   1.060$-$2  &  8.780$-$3  &  5.630$-$3 \\
    1  &   67  &  1.706$-$2  &  1.692$-$2  &  1.032$-$2   &   1.480$-$2  &  1.230$-$2  &  7.860$-$3 \\
    1  &   68  &  1.347$-$3  &  6.543$-$4  &  2.369$-$4   &   2.500$-$3  &  9.020$-$4  &  3.670$-$4 \\
    1  &   69  &  1.615$-$3  &  8.891$-$4  &  3.279$-$4   &   3.230$-$3  &  1.160$-$3  &  4.740$-$4 \\
    1  &   70  &  1.979$-$3  &  1.043$-$3  &  3.803$-$4   &   3.940$-$3  &  1.420$-$3  &  5.770$-$4 \\
    1  &   71  &  3.586$-$3  &  1.864$-$3  &  1.147$-$3   &   6.280$-$3  &  2.050$-$3  &  1.100$-$3 \\
    1  &   72  &  3.020$-$3  &  3.098$-$3  &  1.574$-$3   &   3.550$-$3  &  3.010$-$3  &  1.560$-$3 \\
    1  &   73  &  4.243$-$3  &  4.330$-$3  &  2.196$-$3   &   4.980$-$3  &  4.210$-$3  &  2.180$-$3 \\
    1  &   74  &  5.468$-$3  &  5.587$-$3  &  2.843$-$3   &   6.360$-$3  &  5.420$-$3  &  2.800$-$3 \\
    1  &   75  &  2.261$-$2  &  3.845$-$2  &  6.213$-$2   &   1.540$-$2  &  2.090$-$2  &  4.080$-$2 \\
    1  &   76  &  4.259$-$3  &  4.016$-$3  &  5.216$-$3   &   8.410$-$3  &  4.280$-$3  &  5.510$-$3 \\
    1  &   77  &  2.331$-$3  &  5.849$-$3  &  8.618$-$3   &   8.480$-$3  &  3.450$-$3  &  1.850$-$3 \\
    1  &   78  &  8.539$-$3  &  1.309$-$2  &  1.337$-$2   &   9.240$-$3  &  1.300$-$2  &  1.360$-$2 \\
\hline  
\end{tabular}

\begin{flushleft}
{\small
DARC: Present results with the {\sc darc} code \\ 
ICFT: Results of  \cite{icft}   with the {\sc icft} code \\ 
}
\end{flushleft}
\end{table*}

Our calculated values of $\Upsilon$ are listed in Table~8 over a wide temperature range up to 10$^{5.9}$ K, suitable for applications to a wide range of astrophysical (and laboratory) plasmas. Corresponding data at any intermediate temperature can be easily interpolated, because (unlike $\Omega$) $\Upsilon$ is a slowly varying function of T$_e$. As already noted in section~1, the most recent, extensive and perhaps  best available corresponding data for $\Upsilon$ are those by \cite{icft}.  Similar to ourselves, they have adopted the (semi-relativistic) $R$-matrix code,  resolved resonances in a fine energy mesh,  averaged $\Omega$ over a Maxwellian distribution, and  reported results for transitions among 238 levels, over a wide range of electron temperature up to 1.8$\times$10$^7$ K. Therefore, we undertake a  comparison only with their results.

In Table~9  we compare our  results for $\Upsilon$ with those of \cite{icft} at three  temperatures of 1800, 90~000 and 450~000 K, for resonance transitions up to level 78. The first and the third are the lowest and the highest {\em common} temperatures between the two calculations whereas the second is the most appropriate for applications to astrophysical plasmas, as already mentioned in section~5. Comparisons of $\Upsilon$ for all transitions among the lowest 78 levels  at these  three temperatures are shown in Fig.~3 (a, b and c) in the form of the ratio R = $\Upsilon_{DARC}$/$\Upsilon_{ICFT} $.  Note that negative values of R represent $\Upsilon_{ICFT}$/$ \Upsilon_{DARC}$, i.e.  $\Upsilon_{ICFT}  >  \Upsilon_{DARC}$.  At the two higher values of T$_e$, the two sets of $\Upsilon$ generally agree within 20\% for most of the transitions listed in Table~9. However, differences are larger (up to a factor of two) for a few, particularly those with upper levels $>$ 60. Unfortunately, the discrepancies are larger (up to a factor of two) at the lowest temperature of 1800 K, for about half the transitions, and in a majority of cases $\Upsilon_{ICFT} > \Upsilon_{DARC}$. Such discrepancies at low temperatures are not uncommon and mostly arise due to the position of resonances, because 1800 K is equivalent  to only $\sim$0.011 Ryd, whereas our adopted energy mesh is 0.001 Ryd. To reliably calculate $\Upsilon$ at such low temperature(s), the energy mesh needs to be {\em finer} because our tests show that the uncertainty introduced by our (comparatively) coarse mesh is $\sim$20\% at T$_e$ = 1800 K.

\cite{icft} adopted a comparatively finer energy mesh of 0.000~04 Ryd, but only for $J \le$ 11.5 for which they performed the {\em electron-exchange} calculations. For higher partial waves, up to $J$ = 45, they adopted a coarser mesh of 0.004 Ryd in a {\em no-exchange} calculation. These two different meshes should not normally matter for calculations of $\Upsilon$ at low temperatures, such as 1800 K, because   $J \le$ 11.5 should be sufficient for the convergence of $\Omega$ at low energies. Therefore, their reported $\Upsilon$ {\em may be} comparatively more accurate at this T$_e$. For transitions among all levels, in  most cases the $\Upsilon$ of \cite{icft} are larger as shown in Fig.~3a, but in a few instances $ \Upsilon_{DARC} > \Upsilon_{ICFT}$. Examples where our data are larger include  transitions 38--56, 39--56 and 40--56, which are {\em forbidden} and have resonances close to the thresholds. As the calculations of \cite{icft} are primarily in  $LS$ coupling, there is a possibility of these resonances being missed. Nevertheless,  this low temperature of 1800 K is not generally expected to be important for applications in  plasma modelling and therefore we  focus our attention on comparison at the  two higher temperatures. 

As shown in Fig.~3 (b and c) the discrepancies between our calculations and those of \cite{icft}  increase with increasing temperature. For about one fifth of the transitions, the  differences are larger than 20\%, and in a majority of cases $\Upsilon_{ICFT} > \Upsilon_{DARC}$,  by up to a factor of 20. These discrepancies between the two independent calculations are consistent with those already found  for other Be-like ions, namely Al~X, Cl~XIV, K~XVI, Ti~XIX  and Ge~XXIX  \citep{belike}. As  discussed in  \cite{belike}, the main source of inaccuracy in the calculations of \cite{icft} for $\Upsilon$ is the use of a limited energy range for calculating $\Omega$, up to only 11 Ryd which is insufficient for the determination of $\Upsilon$ up to 1.8$\times$10$^7$ K (or equivalently to 114 Ryd), because the integral in Eq. (3) will not converge, although they have extended the energy range of $\Omega$ following the suggested formulae of \cite{bt}. However, in our work  there is no requirement for such an extension  because calculations for $\Omega$ have already  been performed up to sufficiently high energies, as detailed in section~5. Therefore, based on the comparisons discussed here as well as for other Be-like ions and shown in Fig.~3, the  results of $\Upsilon$ by \cite{icft} appear to be overestimated for at least 20\% of the transitions among the lowest 78 levels of C~III.

\section{Conclusions}

Energies  for the lowest 166 levels of  C~III belonging to the $n \le$ 5 configurations have been calculated  with the {\sc grasp} code. For the lowest 10 levels, discrepancies with  measurements are up to 8\%, but agreement is better than 1\% for the remaining 156. Radiative rates are also listed  for four types of transitions (E1, E2, M1 and M2) and no large discrepancies are noted for a majority of strong E1 transitions. These rates have been further employed to calculate lifetimes which are found to be in good agreement, for most  levels, with other theoretical work as well as experimental values. Based on several comparisons, including the velocity and length ratios,  our results for radiative rates, oscillator strengths, line strengths  and lifetimes are judged to be accurate to better than 20\% for a majority of  strong transitions. 

Collision strengths  are reported for resonance transitions, at  energies above thresholds, to facilitate future comparisons as at present no similar data exist. However,  results for the more useful parameter $\Upsilon$  are listed for {\em all} transitions among the 166 levels of C~III and over a wide range of temperature up to 10$^{5.9}$ K. This range of T$_e$ should be sufficient  for the modelling of  a variety of plasmas, such as astrophysical and fusion. Comparisons of $\Upsilon$ have been made with the most recent results of \cite{icft}, which are of comparable complexity and are available for a wider range of transitions and temperatures. For over one fifth of transitions, among the lowest 78 levels, the $\Upsilon$ of \cite{icft} differ with our results by more than 20\% and by up to a factor of 20. In most instances the $\Upsilon$ of \cite{icft} are larger, consistent with the pattern found for other Be-like ions with 13 $\le$ Z $\le$ 32 \citep{belike}.

The accuracy of our $\Omega$ and $\Upsilon$ data is assessed to be better than 20\%, for a majority of transitions, forbidden as well as allowed. This assessment is made partly based on comparisons with available results and our experience with other Be-like ions, and mainly because: (i) we have included a large range of partial waves  to achieve convergence of $\Omega$ at all energies, (ii) have included the contribution of higher neglected partial waves through a top-up, (iii) have  resolved resonances in a fine energy mesh to account for their contribution, and more importantly (iv) have calculated $\Omega$ over a wide range of energy which allowed us to determine $\Upsilon$ up to the highest temperature of our calculation, without any extrapolation.  Hence, we see no  apparent limitations in our  data and hope these  can be confidently applied to the modelling of plasmas. Nevertheless, scope remains for improvement, as for any calculation. This can perhaps be achieved  by the inclusion of  levels of the configurations with  $n >$  5 in the generation of wave functions and the scattering process. 

\section*{Acknowledgments}

KMA is thankful to  AWE Aldermaston for    financial support.

\end{document}